\documentclass[aps,prl,twocolumn,superscriptaddress]{revtex4-1}

\usepackage{graphicx}
\usepackage{color}
\usepackage{siunitx}
\usepackage[english]{babel}
\usepackage{amsmath}

\begin{document}

\title{Scanning Gate Microscopy in a Viscous Electron Fluid}

\author{B.~A.~Braem}
\email{bbraem@phys.ethz.ch}
\affiliation{ETH Z\"urich, Solid State Physics Laboratory, Otto-Stern-Weg 1, 8093 Z\"urich, Switzerland}
\author{F.~M.~D.~Pellegrino}
\affiliation{Dipartimento di Fisica e Astronomia, Universit\`a di Catania, Via S.
Sofia, 64, I-95123 Catania, Italy}
\affiliation{INFN, Sez. Catania, I-95123 Catania, Italy}
\author{A.~Principi}
\affiliation{School of Physics and Astronomy, University of Manchester, Manchester, M13 9PL, United Kingdom}
\author{M.~R\"o\"osli}
\affiliation{ETH Z\"urich, Solid State Physics Laboratory, Otto-Stern-Weg 1, 8093 Z\"urich, Switzerland}
\author{C.~Gold}
\affiliation{ETH Z\"urich, Solid State Physics Laboratory, Otto-Stern-Weg 1, 8093 Z\"urich, Switzerland}
\author{S.~Hennel}
\affiliation{ETH Z\"urich, Solid State Physics Laboratory, Otto-Stern-Weg 1, 8093 Z\"urich, Switzerland}
\author{J.~V.~Koski}
\affiliation{ETH Z\"urich, Solid State Physics Laboratory, Otto-Stern-Weg 1, 8093 Z\"urich, Switzerland}
\author{M.~Berl}
\affiliation{ETH Z\"urich, Solid State Physics Laboratory, Otto-Stern-Weg 1, 8093 Z\"urich, Switzerland}
\author{W.~Dietsche}
\affiliation{ETH Z\"urich, Solid State Physics Laboratory, Otto-Stern-Weg 1, 8093 Z\"urich, Switzerland}
\author{W.~Wegscheider}
\affiliation{ETH Z\"urich, Solid State Physics Laboratory, Otto-Stern-Weg 1, 8093 Z\"urich, Switzerland}
\author{M.~Polini}
\affiliation{Istituto Italiano di Tecnologia, Graphene Labs, Via Morego 30, I-16163 Genova, Italy}
\author{T.~Ihn}
\affiliation{ETH Z\"urich, Solid State Physics Laboratory, Otto-Stern-Weg 1, 8093 Z\"urich, Switzerland}
\author{K.~Ensslin}
\affiliation{ETH Z\"urich, Solid State Physics Laboratory, Otto-Stern-Weg 1, 8093 Z\"urich, Switzerland}


\date{\today}


\begin{abstract}
We measure transport through a Ga[Al]As heterostructure at temperatures between \SI{32}{mK} and \SI{30}{K}. Increasing the temperature enhances the electron-electron scattering rate and viscous effects in the two-dimensional electron gas arise. To probe this regime we measure so-called vicinity voltages and use a voltage-biased scanning tip to induce a movable local perturbation. 
We find that the scanning gate images differentiate reliably between the different regimes of electron transport.
Our data are in good agreement with recent theories for interacting electron liquids in the ballistic and viscous regimes stimulated by measurements in graphene. However, the range of temperatures and densities where viscous effects are observable in Ga[Al]As are very distinct from the graphene material system.
\end{abstract}

\maketitle

Inter-particle collisions dominate the behavior of fluids as described by hydrodynamic theory \cite{landau_fluid_1987}. In degenerate, clean two-dimensional electron gases (2DEGs), e.g. realized in Ga[Al]As heterostructures or in graphene, hydrodynamic behavior may be expected if electron-electron interaction is the dominant scattering mechanism. At millikelvin temperatures, however, electron-impurity scattering dominates over electron-electron scattering. 
The latter produces only small corrections accounted for within Fermi-liquid theory, a description involving weakly interacting quasiparticles. 
The relevance of electron-electron scattering is enhanced by increasing the temperature, thus softening the Fermi surface. The electron-electron scattering length $l_\mathrm{ee}$ then reaches well below both the geometric device sizes and the momentum relaxation length.
Early experiments realized this regime aiming at the identification of hydrodynamic effects in Ga[Al]As 2DEGs \cite{molenkamp_observation_1994, de_jong_hydrodynamic_1995}.  
Very recently, experimental signatures of viscosity due to electron-electron interaction have been found in graphene \cite{bandurin_negative_2016, crossno_observation_2016}, Ga[Al]As \cite{gusev_viscous_2018}, PdCoO$_2$ \cite{moll_evidence_2016}, and WP$_2$ \cite{gooth_electrical_2017}, and related theories have been developed \cite{torre_nonlocal_2015,levitov_electron_2016,pellegrino_electron_2016, bandurin_probing_2018,shytov_electron_2018}.

Viscous flow gives rise to intricate spatial flow patterns occurring at length scales well below the Drude scattering length $l_\mathrm{D}$, beyond which the momentum of the electronic system is dispersed \cite{torre_nonlocal_2015,levitov_electron_2016, pellegrino_electron_2016}. 
Such spatial patterns in electronic systems have been theoretically predicted, but so far not been imaged experimentally. This motivates us to perform scanning gate microscopy \cite{eriksson_cryogenic_1996, topinka_imaging_2000} measurements on a 2DEG in a Ga[Al]As heterostructures with signatures of viscous charge carrier flow.
We find that the scanning gate measurement distinguishes the ballistic and viscous regimes of transport with high sensitivity. In the viscous regime, the scanning tip can locally revive ballistic contributions to the measured signals by introducing new and tunable length scales to the system geometry. Both a hydrodynamic and a ballistic model of electron transport guide us in interpreting the experimental data.

\begin{figure}
 \includegraphics[width=\linewidth]{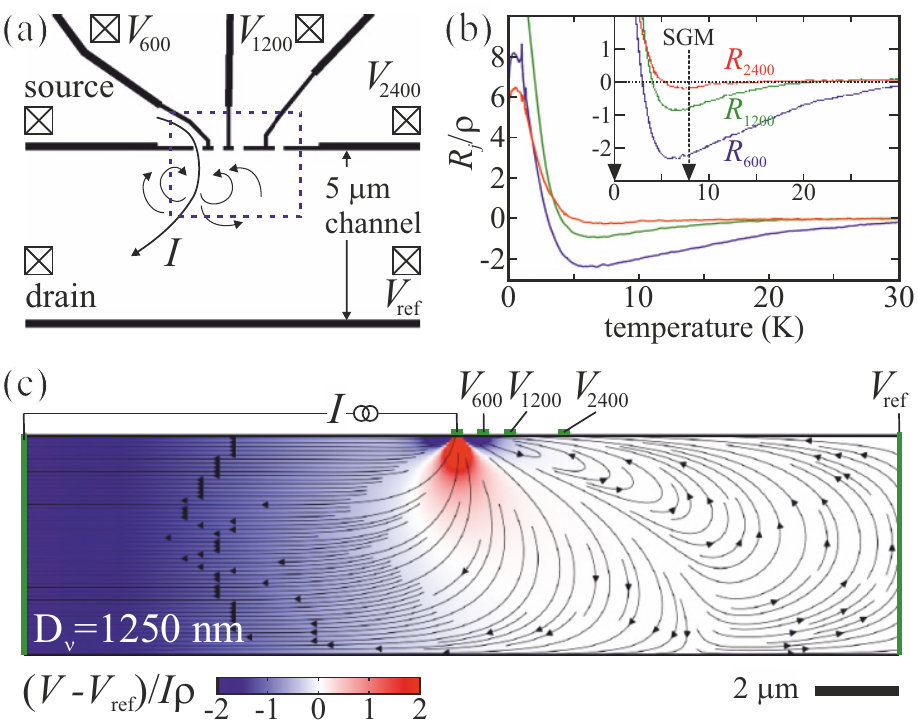}
 \caption{(Color online) (a) Top-gates (indicated by black lines) deplete the 2DEG to shape the sample to a channel with orifices to the top region, which serve as current injector and voltage probes. The vicinity voltages $V_j$ are measured with respect to the channel potential $V_\mathrm{ref}$. 
 Arrows indicate schematically the current distribution if back-flow occurs due to viscosity. The dashed rectangle marks the area where the tip of the scanning gate microscope is scanned. (b) Normalized vicinity resistances $R_j/\rho:=(V_j-V_\mathrm{ref})/I\rho$ as a function of temperature in the absence of the SGM tip at $n=\SI{1.2e11}{\per\square\centi\meter}$. The inset shows the same data enlarged to highlight the minima at around \SI{7}{K}. The vertical dashed lines mark the temperatures of the SGM measurements in Fig.~\ref{fig2label}. (c) Current distribution and potential from solving the hydrodynamic model with a length scale parameter $D_\nu=\SI{1.25}{\micro\meter}$, which corresponds to $n=\SI{1.2e11}{\per\square\centi\meter}$ and $T \approx\SI{7}{K}$. The green lines mark equipotential surfaces forming the contacts to the channel. 
 }
 \label{fig1label}
\end{figure}
 
Following the experiments by Bandurin \textit{et al.} \cite{bandurin_negative_2016, bandurin_probing_2018} on graphene, we use vicinity voltage probes close to a local current injector to measure effects of viscosity. The concept of the measurement is sketched in Fig.~\ref{fig1label}(a). 
We pass a current $I$ from the source contact through a \SI{300}{nm} wide orifice into a \SI{5}{\micro\meter} wide channel, which is connected to the drain contact at ground potential.
The upper channel boundary has three additional openings to probe the vicinity voltages $V_j$ at a distance $d_j$ from the current-injecting orifice with $d_j$ being \SI{600}{\nano\meter}, \SI{1200}{\nano\meter}, and \SI{2400}{\nano\meter} respectively. 
The vicinity voltages $V_j$ are measured with respect to the reference potential $V_\mathrm{ref}$ at the right end of the channel. In this geometry one expects positive vicinity voltages for diffusive and ballistic electron motion in the channel, and negative values if electron-electron interaction is dominant \cite{torre_nonlocal_2015, bandurin_negative_2016, shytov_electron_2018}. In the latter case back-flow currents are proposed \cite{pellegrino_electron_2016} as indicated by the schematic flow pattern in Fig.~\ref{fig1label}(a). 

We use a Ga[Al]As heterostructure with a 2DEG buried 130~nm below the surface and a back-gate to tune the electron density $n$ \cite{berl_structured_2016}.
The supplemental material provides experimental details, e.g. measurement parameters, and electron density as a function of back-gate voltage. Applying negative voltages to the top-gates defines the structure shown in Fig.~\ref{fig1label}(a) by locally depleting the 2DEG.
To measure the vicinity voltages we use low-noise voltage amplifiers and standard lock-in techniques at 31.4~Hz. We cool the sample in a cryostat equipped with an atomic force microscope to create a local perturbation by scanning gate microscopy (SGM). 

We define the vicinity resistance as the ratio $R_j=V_j/I$ of the measured quantities, without offset-subtraction. Figure~\ref{fig1label}(b) shows the vicinity resistances normalized to the 2DEG sheet resistance $\rho$ as a function of temperature $T$ from 30~mK to 30~K. At the lowest temperature, all vicinity resistances are positive. With increasing $T$ their signs change at around \SI{3}{K}. The temperature of the zero-crossing increases with $d_j$. Furthermore, the vicinity resistances have a minimum at around 7~K and tend towards zero with increasing $T$. This behavior is similar to recent experiments in bilayer graphene \cite{bandurin_negative_2016, bandurin_probing_2018}.

\begin{figure}
 \includegraphics[width=\linewidth]{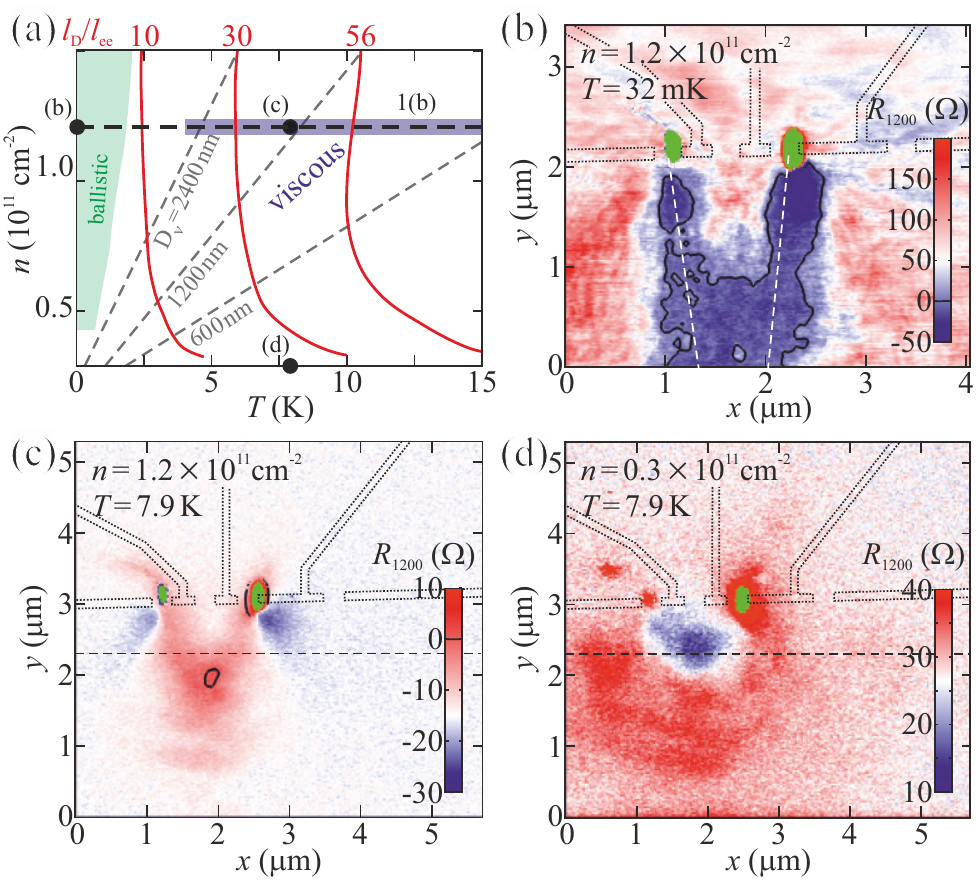}
 \caption{(a) Schematic of transport regimes as a function of temperature and electron density. Viscous effects are expected at a high ratio $l_\mathrm{D}/l_\mathrm{ee}$. Red lines mark contours of $l_\mathrm{D}/l_\mathrm{ee}$ and show the increase with $T$ and $n$. 
The green shade marks the ballistic regime where both $l_\mathrm{D}$ and $l_\mathrm{ee}$ exceed the channel width. Dashed grey lines indicate $D_\nu=d_j$.
Three black dots mark the parameters of the SGM measurements in panels (b)-(d). The data shown in Fig.~\ref{fig1label}(b) is measured along the dashed black line, the blue shade indicates the temperature range of negative $R_{1200}$. 
(b)-(d) Vicinity resistance $R_{1200}$ as a function of SGM tip position $x,y$ with white color marking the value in the absence of the tip: (b) At 32~mK we observe a V-shape of reduced $R_{1200}(x,y)$ along the white dashed lines, which mark the ballistic trajectory. Dotted lines mark the outlines of the gates, areas of green color indicate tip positions leading to $I=0$ or disconnected voltage probe.
 (c) At 7.9~K the vicinity resistance $R_{1200}(x,y)$ shows a maximum instead of the~V. (d) $R_{1200}(x,y)$ at 7.9~K at lower electron density.
 }
 \label{fig2label}
\end{figure}

To understand the behavior of the vicinity resistances as a function of temperature in Fig.~\ref{fig1label}(b) we consider the scattering lengths $l_\mathrm{ee}$ and $l_\mathrm{D}$ of the 2DEG realized within the range of our experimental parameters. Figure~\ref{fig2label}(a) displays red contour lines of the ratio $l_\mathrm{D}/l_\mathrm{ee}$, where $l_\mathrm{ee}=v_\mathrm{F} \tau_\mathrm{ee}$ was calculated from $\tau_\mathrm{ee}$ \cite{jungwirth_electron-electron_1996,giuliani_quantum_2008} and the Drude scattering length $l_\mathrm{D}$ was extracted from bulk resistance measurements (absolute values of $l_\mathrm{ee}$ and $l_\mathrm{D}$ in supplemental material). One can see that $l_\mathrm{D}/l_\mathrm{ee} \gg 1$ in an extended region of the parameter space indicating where electron-electron interactions dominate. The horizontal dashed line marks the density of the measurement shown in Fig.~\ref{fig1label}(b). 
Two complementary theories exist describing the behavior along this line. Their applicability depends on the ratio $l_\mathrm{ee}/d_j$. 

The regime $l_\mathrm{ee} < d_j$ realized for $T \gtrsim \SI{6}{K}$ is described by the viscous theory \cite{torre_nonlocal_2015, bandurin_negative_2016, levitov_electron_2016}. Numerical calculations as in Ref.~\citenum{bandurin_negative_2016} based on the solution of the Navier-Stokes equation result in the flow patterns shown in Fig.~\ref{fig1label}(c) for our sample geometry. The intrinsic length scale of the theory $D_\nu=\sqrt{l_\mathrm{ee}l_\mathrm{D}/4}$ was chosen to match the experimental conditions at about \SI{7}{K}.
The theory predicts negative vicinity resistances of $R_{600}/\rho=-0.65$, $R_{1200}/\rho=-0.11$, and $R_{2400}/\rho=-0.015$, which are in qualitative agreement with the measurements in Fig.~\ref{fig1label}(b). With increasing temperature or $d_j$, $D_\nu$ falls below $d_j$ and the vicinity voltage probes become insensitive to the quasi-local viscous effects. This is in accordance with $R_j/\rho$ in Fig.~\ref{fig1label}(b) tending towards zero for high $T$.

For $l_\mathrm{ee} > d_j$, i.e. $T \lesssim \SI{4}{K}$, diffusive transport between the injector and the voltage probe is not effective yet, and single electron-electron scattering events will dominate the measured vicinity voltages. This regime is described by the theory of Shytov \textit{et al.} \cite{shytov_electron_2018}. They propose that the vicinity voltage response is negative with its strength increasing with the electron-electron scattering rate, i.e. with temperature. This is in qualitative agreement with the strongly decreasing $R_j$ around \SI{3}{K} in Fig.~\ref{fig1label}(b).

At temperatures below \SI{1.7}{K}, $l_\mathrm{ee}$ exceeds the width of the channel of our sample and both of the above mentioned theories become inapplicable. 
An extended theory covering the full range of temperatures 
\cite{bandurin_probing_2018} proposes that the positive vicinity voltage observed in the experiment is caused by ballistic electron motion between the injector orifice and the voltage probe with intermittent reflection at the opposite channel boundary. This claim is supported by the SGM measurements presented below.

We now scan the SGM tip at a fixed height of \SI{40}{nm} above the GaAs surface in the area indicated by the dashed rectangle in Fig.~\ref{fig1label}(a). 
Applying a negative voltage to the tip creates a disk of depleted 2DEG with a diameter of approximately \SI{300}{nm}. 
We have taken scanning gate images for a range of back-gate voltages, contact configurations and channel widths, but in the interest of brevity we present data for the three selected, most significant regimes marked by the black dots in Fig.~\ref{fig2label}(a). 

Figure~\ref{fig2label}(b) shows the vicinity resistance $R_{\mathrm{1200}}$ as a function of the tip position $x,y$ at $T=32$~mK, in the ballistic regime where $l_\mathrm{D}\approx \SI{36}{\micro\meter}$ and $l_\mathrm{ee}\gg l_\mathrm{D}$. 
White color presents $R_{\mathrm{1200}}$ as measured in the absence of the tip. Blue indicates a reduced, and red an increased value of $R_{\mathrm{1200}}$. The black contour at zero highlights the tip positions of sign inversion.
For orientation, black dotted lines mark the outlines of the top-gates. If the tip depletes the 2DEG in the source orifice or in the voltage probe opening, $R_{\mathrm{1200}}$ cannot be extracted and the position is colored green. 
The classical ballistic electron trajectory from the source to the voltage probe, that is once reflected by the channel gate, is indicated by white dashed lines. We observe a V-shaped reduction of $R_{\mathrm{1200}}$ along the outline of this ballistic path.
We interpret the result in the following way: In the absence of the tip, some electrons are ballistically reflected by the channel gate into the voltage probe and we measure positive $R_{1200}$. 
For tip positions along the V-shaped ballistic path, the tip potential deflects ballistic trajectories and we observe a reduction of $R_{1200}(x,y)$. Conversely, a tip positioned outside the V guides additional trajectories into the voltage probe and thus increases $R_{1200}(x,y)$. Such a deflection of ballistic trajectories has been demonstrated by earlier SGM work \cite{crook_imaging_2000,aidala_imaging_2007,bhandari_imaging_2016}.

We change to the viscous regime by heating the cryostat temperature to 7.9~K such that $l_\mathrm{D}\approx\SI{16}{\micro\meter}$ and $l_\mathrm{ee}\approx\SI{0.4}{\micro\meter}<d_j$, leading to a characteristic length scale $D_\nu=\SI{1.2}{\micro\meter}$. Figure~\ref{fig2label}(c) shows the corresponding SGM measurement. The striking difference to Fig.~\ref{fig2label}(b) witnesses the change of the transport regime from ballistic to viscous. The V-shaped reduction of $R_{1200}$ is no longer present. Consistent with the measurements in Fig.~\ref{fig1label}(b), $R_{\mathrm{1200}}(x,y)$ is negative if the tip is far from source orifice or voltage probe, for example at $x > \SI{5}{\micro\meter}$. In contrast to measurements at lower temperature, $R_{\mathrm{1200}}(x,y)$ features a maximum at $x \approx y \approx \SI{2}{\micro\meter}$. This distinguished position is approximately separated by $d_{1200}$ from both the source orifice and the voltage probe. Here the tip forms a scattering site much closer than the lower channel edge at $y \approx \SI{-2}{\micro\meter}$.

We now reduce the electron density to $n=\SI{0.3e11}{\per\square\centi\meter}$ while keeping the temperature at 7.9 K (see the point labeled (d) in Fig.~\ref{fig2label}(a).) 
At this low density, $l_\mathrm{D}\approx\SI{1.6}{\micro\meter}$ and $l_\mathrm{ee}\approx\SI{70}{nm} \ll d_j$, and the characteristic scale $D_\nu = \SI{170}{nm}$ has fallen well below $d_j$. 
Therefore we do not observe the effects of viscosity but a positive vicinity resistance in the absence of the tip. 
SGM at this low density finds $R_{\mathrm{1200}}(x,y)$ presented in Fig.~\ref{fig2label}(d), which is significantly different to both the result in (b) and (c) at four times higher electron density.
Instead of a maximum we find a $R_{\mathrm{1200}}(x,y)$  minimum at $x \approx \SI{2}{\micro\meter},y \approx \SI{2.3}{\micro\meter}$.

\begin{figure}
 \includegraphics[width=\linewidth]{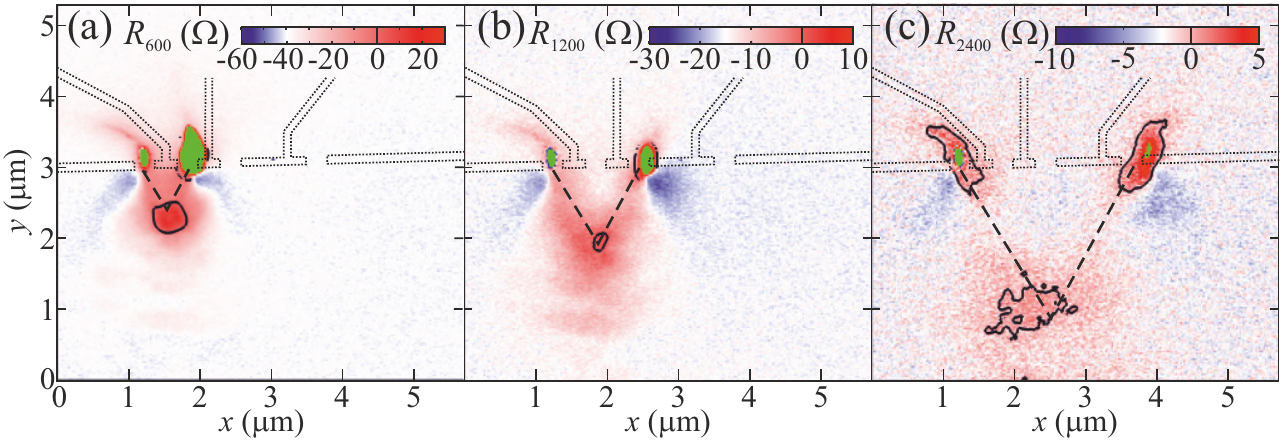}
 \caption{All three vicinity resistances at $T=\SI{7.9}{\kelvin}$ and $n=\SI{1.2e11}{\per\square\centi\meter}$ as a function of tip position: (a) $R_{600}$, (b) $R_{1200}$ as already shown in Fig.~\ref{fig2label}(c), and (d) $R_{2400}$. As indicated by the dashed lines, we find a maximum of $R_j$ when the tip forms an equilateral triangle with the source orifice and the voltage probe. 
 }
 \label{fig3label}
\end{figure}
In Fig.~\ref{fig3label} we return to the high-density regime and compare all three vicinity resistances $R_j$ measured at \SI{7.9}{K}. Note that Fig.~\ref{fig3label}(b) reproduces Fig.~\ref{fig2label}(c) for convenience. 
The dashed lines form an equilateral triangle between the current-injecting orifice and the respective vicinity voltage probe. The tip of the triangle coincides with the maximum of $R_j$ in all three images, 
suggesting a purely geometrical interpretation. 
It seems that the presence of the tip-induced potential in this symmetry point prevents the observation of viscous effects and reestablishes a positive vicinity voltage.

In conjunction with Figs.~\ref{fig1label}(b) and \ref{fig2label}(a) we have already discussed the microscopic transport regimes which we now found to result in dramatic differences in the scanning gate images in Figs.~\ref{fig2label}(b)-(d). 
In the remaining parts of the paper, we discuss the imaging mechanism of the scanning gate technique in the viscous regime represented by Figs.~\ref{fig2label}(c) and (d).
Naively one could think that the scanning tip-induced potential introduces a new internal sample boundary, which leads to a reorganization of the viscous flow pattern and thereby to a change in the vicinity voltages. We will therefore discuss the agreement and differences between the hydrodynamic model in Fig.~\ref{fig1label}(c) and the scanning gate measurements first.

\begin{figure}
 \includegraphics[width=\linewidth]{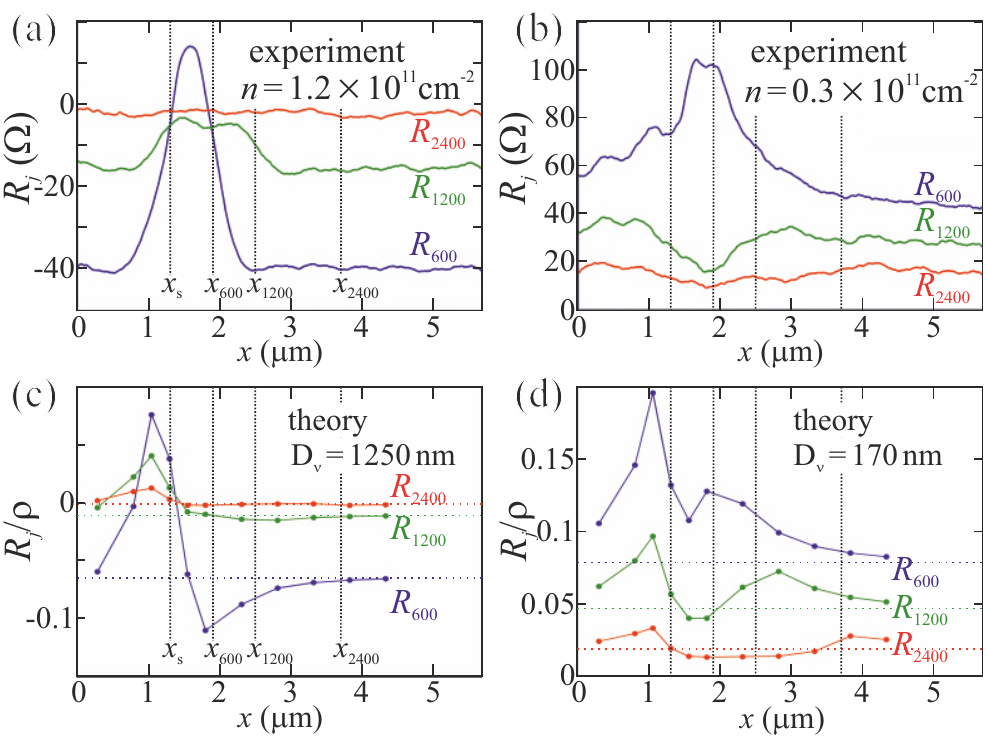}
 \caption{Comparison between experiment and hydrodynamical model: (a), (b) $R_j$ along the dashed lines in Figs.~\ref{fig2label}(c) and (d), the $x$-coordinates $x_j$ of source orifice and voltage probes are marked by the vertical lines. (c), (d) Vicinity resistances calculated with the hydrodynamic model for the tip positions and length scales $D_\nu$ in the experimental data of (a) and (b). The horizontal dotted lines denote the vicinity resistances in the absence of the tip.
 }
 \label{fig4label}
\end{figure}
The hydrodynamic model solves for the stationary flow of the classical incompressible viscous electron liquid at very low Reynolds numbers, where the non-linear convective acceleration term in the Navier-Stokes equation can be neglected. Thanks to the addition of a Drude-like momentum relaxation rate, the resulting equations are well suited to describe the transition from the viscous to the momentum-scattering dominated regime \cite{torre_nonlocal_2015}. However, this model does not account for ballistic effects.
We solve the model in the presence of a local Lorentzian-shaped decrease of the electron density caused by the tip potential \cite{eriksson_effect_1996} (details of the tip implementation in supplemental material). 

In Fig.~\ref{fig4label} we compare the measured vicinity resistances along the dashed lines in Figs.~\ref{fig2label}(c) and (d) with the prediction of the model for the same tip positions and length scales $D_\nu$. For orientation, the vertical lines mark the $x$-coordinates of the source orifice and the voltage probes. In the high-density case in (a), (c) we find qualitative agreement for tip positions $x>\SI{4}{\micro\meter}$, but not at $x<\SI{3}{\micro\meter}$ where the distance between the tip and the orifices is of the order of $D_\nu$ and no longer $\gg l_\mathrm{ee}$.
We speculate that the disagreement originates from the close, tip-induced scattering site which revives ballistic effects. 

In the low-density case (b), (d) we find a rough agreement for all tip positions for $R_{1200}$ and $R_{2400}$, but not for the signal $R_{600}$ if the tip is close to the respective voltage probe. 
As in the high-density case, we find a disagreement if the distance between the tip and the orifices is of the order of $D_\nu$.
Since the hydrodynamic model does not describe ballistic effects, we consider this as a justification for the hypothesis, that the presence of the tip leads to a revival of ballistic effects in the sample on the small length scale introduced by the tip.

\begin{figure}
 \includegraphics[width=\linewidth]{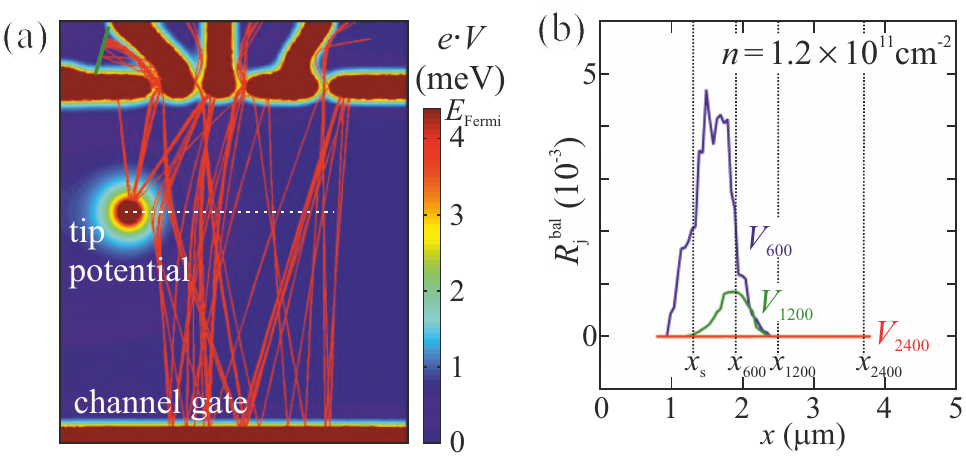}
 \caption{Classical trajectories: (a) Color plot showing the potential landscape in the 2DEG from tip and top-gates from finite element simulation. Red lines show classical trajectories starting at the green line in the source lead and ending in one of the voltage probes. (b) The number of trajectories ending in the voltage probes weighted by the trajectory length. 
 }
 \label{fig5label}
\end{figure}
To test this hypothesis, we investigate ballistic contributions in a deliberately oversimplified classical model. We calculate electron trajectories emanating from the source orifice in the electrostatic potential of gates and tip exemplarily shown in Fig.~\ref{fig5label}(a). For tip positions along the dashed line we count the number of trajectories that end in one of the voltage probes as a qualitative measure for the ballistic contribution $R_j^\mathrm{bal}$ to the corresponding vicinity resistance. We count each trajectory with a weight that decreases exponentially with trajectory length to account for electron-electron scattering (details in supplemental material). 
Figure~\ref{fig5label}(b) shows the resulting maxima of $R_j^\mathrm{bal}$ for the tip positions in the middle between the source orifice and the corresponding voltage probe. This is in agreement with the experimental observations at high density in Fig.~\ref{fig4label}(a), when the tip is close to the orifices. It supports our speculative interpretation that the resistance maxima in Fig.~\ref{fig3label} result from an enhancement of ballistic contributions to the conductance, which quench the visibility of the viscous effects.

In summary, we have presented measurements of negative vicinity resistances in Ga[Al]As heterostructures, which indicate viscous behavior. By increasing the temperature we observed the transition from the ballistic to the viscous regime when the electron-electron scattering length falls below the separation between current injector and voltage probes. 
These findings are qualitatively similar to observations on graphene samples, but both the charge carrier density and the characteristic temperature are an order of magnitude lower.
The movable perturbation by SGM introduces an additional, competing length scale. Scanning gate images in the ballistic and viscous regimes are markedly different. By forming a scattering site close to the source orifice and the voltage probes, ballistic effects can be restored even though the electron-electron scattering length is below the channel width.
A hydrodynamic model explains some of the observed features including the negative vicinity resistances. From the difference between this model and the experiment we find that residual ballistic effects need to be considered on small length scales even at a high temperature of \SI{7.9}{\kelvin}. The theory developed in Ref.~\cite{bandurin_probing_2018} based on the kinetic equation is well suited to describe the transition between the ballistic and the viscous regime of transport. 
It therefore remains an interesting open question, if this approach could be used for describing the scanning gate experiment, and if it yields agreement with the experiment over a larger range of parameters.

\begin{acknowledgments}
We thank Leonid Levitov and Yigal Meir for valuable discussions. 
The authors acknowledge financial support from ETH Z\"urich and from the Swiss National Science Foundation (NCCR QSIT, SNF 2-77255).
\end{acknowledgments}

\bibliography{mybib}

\begin{thebibliography}{25}%
\makeatletter
\providecommand \@ifxundefined [1]{%
 \@ifx{#1\undefined}
}%
\providecommand \@ifnum [1]{%
 \ifnum #1\expandafter \@firstoftwo
 \else \expandafter \@secondoftwo
 \fi
}%
\providecommand \@ifx [1]{%
 \ifx #1\expandafter \@firstoftwo
 \else \expandafter \@secondoftwo
 \fi
}%
\providecommand \natexlab [1]{#1}%
\providecommand \enquote  [1]{``#1''}%
\providecommand \bibnamefont  [1]{#1}%
\providecommand \bibfnamefont [1]{#1}%
\providecommand \citenamefont [1]{#1}%
\providecommand \href@noop [0]{\@secondoftwo}%
\providecommand \href [0]{\begingroup \@sanitize@url \@href}%
\providecommand \@href[1]{\@@startlink{#1}\@@href}%
\providecommand \@@href[1]{\endgroup#1\@@endlink}%
\providecommand \@sanitize@url [0]{\catcode `\\12\catcode `\$12\catcode
  `\&12\catcode `\#12\catcode `\^12\catcode `\_12\catcode `\%12\relax}%
\providecommand \@@startlink[1]{}%
\providecommand \@@endlink[0]{}%
\providecommand \url  [0]{\begingroup\@sanitize@url \@url }%
\providecommand \@url [1]{\endgroup\@href {#1}{\urlprefix }}%
\providecommand \urlprefix  [0]{URL }%
\providecommand \Eprint [0]{\href }%
\providecommand \doibase [0]{http://dx.doi.org/}%
\providecommand \selectlanguage [0]{\@gobble}%
\providecommand \bibinfo  [0]{\@secondoftwo}%
\providecommand \bibfield  [0]{\@secondoftwo}%
\providecommand \translation [1]{[#1]}%
\providecommand \BibitemOpen [0]{}%
\providecommand \bibitemStop [0]{}%
\providecommand \bibitemNoStop [0]{.\EOS\space}%
\providecommand \EOS [0]{\spacefactor3000\relax}%
\providecommand \BibitemShut  [1]{\csname bibitem#1\endcsname}%
\let\auto@bib@innerbib\@empty
\bibitem [{\citenamefont {Landau}\ and\ \citenamefont
  {Lifshitz}(1987)}]{landau_fluid_1987}%
  \BibitemOpen
  \bibfield  {author} {\bibinfo {author} {\bibfnamefont {L.~D.}\ \bibnamefont
  {Landau}}\ and\ \bibinfo {author} {\bibfnamefont {E.~M.}\ \bibnamefont
  {Lifshitz}},\ }\href@noop {} {\emph {\bibinfo {title} {Fluid {Mechanics},
  {Second} {Edition}: {Volume} 6}}},\ \bibinfo {edition} {2nd}\ ed.\ (\bibinfo
  {publisher} {Butterworth-Heinemann},\ \bibinfo {address} {Amsterdam u.a},\
  \bibinfo {year} {1987})\BibitemShut {NoStop}%
\bibitem [{\citenamefont {Molenkamp}\ and\ \citenamefont
  {de~Jong}(1994)}]{molenkamp_observation_1994}%
  \BibitemOpen
  \bibfield  {author} {\bibinfo {author} {\bibfnamefont {L.~W.}\ \bibnamefont
  {Molenkamp}}\ and\ \bibinfo {author} {\bibfnamefont {M.~J.~M.}\ \bibnamefont
  {de~Jong}},\ }\href {\doibase 10.1016/0038-1101(94)90244-5} {\bibfield
  {journal} {\bibinfo  {journal} {Solid-State Electronics}\ }\textbf {\bibinfo
  {volume} {37}},\ \bibinfo {pages} {551} (\bibinfo {year} {1994})}\BibitemShut
  {NoStop}%
\bibitem [{\citenamefont {de~Jong}\ and\ \citenamefont
  {Molenkamp}(1995)}]{de_jong_hydrodynamic_1995}%
  \BibitemOpen
  \bibfield  {author} {\bibinfo {author} {\bibfnamefont {M.~J.~M.}\
  \bibnamefont {de~Jong}}\ and\ \bibinfo {author} {\bibfnamefont {L.~W.}\
  \bibnamefont {Molenkamp}},\ }\href {\doibase 10.1103/PhysRevB.51.13389}
  {\bibfield  {journal} {\bibinfo  {journal} {Phys. Rev. B}\ }\textbf {\bibinfo
  {volume} {51}},\ \bibinfo {pages} {13389} (\bibinfo {year}
  {1995})}\BibitemShut {NoStop}%
\bibitem [{\citenamefont {Bandurin}\ \emph {et~al.}(2016)\citenamefont
  {Bandurin}, \citenamefont {Torre}, \citenamefont {Kumar}, \citenamefont
  {Shalom}, \citenamefont {Tomadin}, \citenamefont {Principi}, \citenamefont
  {Auton}, \citenamefont {Khestanova}, \citenamefont {Novoselov}, \citenamefont
  {Grigorieva}, \citenamefont {Ponomarenko}, \citenamefont {Geim},\ and\
  \citenamefont {Polini}}]{bandurin_negative_2016}%
  \BibitemOpen
  \bibfield  {author} {\bibinfo {author} {\bibfnamefont {D.~A.}\ \bibnamefont
  {Bandurin}}, \bibinfo {author} {\bibfnamefont {I.}~\bibnamefont {Torre}},
  \bibinfo {author} {\bibfnamefont {R.~K.}\ \bibnamefont {Kumar}}, \bibinfo
  {author} {\bibfnamefont {M.~B.}\ \bibnamefont {Shalom}}, \bibinfo {author}
  {\bibfnamefont {A.}~\bibnamefont {Tomadin}}, \bibinfo {author} {\bibfnamefont
  {A.}~\bibnamefont {Principi}}, \bibinfo {author} {\bibfnamefont {G.~H.}\
  \bibnamefont {Auton}}, \bibinfo {author} {\bibfnamefont {E.}~\bibnamefont
  {Khestanova}}, \bibinfo {author} {\bibfnamefont {K.~S.}\ \bibnamefont
  {Novoselov}}, \bibinfo {author} {\bibfnamefont {I.~V.}\ \bibnamefont
  {Grigorieva}}, \bibinfo {author} {\bibfnamefont {L.~A.}\ \bibnamefont
  {Ponomarenko}}, \bibinfo {author} {\bibfnamefont {A.~K.}\ \bibnamefont
  {Geim}}, \ and\ \bibinfo {author} {\bibfnamefont {M.}~\bibnamefont
  {Polini}},\ }\href {\doibase 10.1126/science.aad0201} {\bibfield  {journal}
  {\bibinfo  {journal} {Science}\ }\textbf {\bibinfo {volume} {351}},\ \bibinfo
  {pages} {1055} (\bibinfo {year} {2016})}\BibitemShut {NoStop}%
\bibitem [{\citenamefont {Crossno}\ \emph {et~al.}(2016)\citenamefont
  {Crossno}, \citenamefont {Shi}, \citenamefont {Wang}, \citenamefont {Liu},
  \citenamefont {Harzheim}, \citenamefont {Lucas}, \citenamefont {Sachdev},
  \citenamefont {Kim}, \citenamefont {Taniguchi}, \citenamefont {Watanabe},
  \citenamefont {Ohki},\ and\ \citenamefont {Fong}}]{crossno_observation_2016}%
  \BibitemOpen
  \bibfield  {author} {\bibinfo {author} {\bibfnamefont {J.}~\bibnamefont
  {Crossno}}, \bibinfo {author} {\bibfnamefont {J.~K.}\ \bibnamefont {Shi}},
  \bibinfo {author} {\bibfnamefont {K.}~\bibnamefont {Wang}}, \bibinfo {author}
  {\bibfnamefont {X.}~\bibnamefont {Liu}}, \bibinfo {author} {\bibfnamefont
  {A.}~\bibnamefont {Harzheim}}, \bibinfo {author} {\bibfnamefont
  {A.}~\bibnamefont {Lucas}}, \bibinfo {author} {\bibfnamefont
  {S.}~\bibnamefont {Sachdev}}, \bibinfo {author} {\bibfnamefont
  {P.}~\bibnamefont {Kim}}, \bibinfo {author} {\bibfnamefont {T.}~\bibnamefont
  {Taniguchi}}, \bibinfo {author} {\bibfnamefont {K.}~\bibnamefont {Watanabe}},
  \bibinfo {author} {\bibfnamefont {T.~A.}\ \bibnamefont {Ohki}}, \ and\
  \bibinfo {author} {\bibfnamefont {K.~C.}\ \bibnamefont {Fong}},\ }\href
  {\doibase 10.1126/science.aad0343} {\bibfield  {journal} {\bibinfo  {journal}
  {Science}\ }\textbf {\bibinfo {volume} {351}},\ \bibinfo {pages} {1058}
  (\bibinfo {year} {2016})}\BibitemShut {NoStop}%
\bibitem [{\citenamefont {Gusev}\ \emph {et~al.}(2018)\citenamefont {Gusev},
  \citenamefont {Levin}, \citenamefont {Levinson},\ and\ \citenamefont
  {Bakarov}}]{gusev_viscous_2018}%
  \BibitemOpen
  \bibfield  {author} {\bibinfo {author} {\bibfnamefont {G.~M.}\ \bibnamefont
  {Gusev}}, \bibinfo {author} {\bibfnamefont {A.~D.}\ \bibnamefont {Levin}},
  \bibinfo {author} {\bibfnamefont {E.~V.}\ \bibnamefont {Levinson}}, \ and\
  \bibinfo {author} {\bibfnamefont {A.~K.}\ \bibnamefont {Bakarov}},\ }\href
  {\doibase 10.1063/1.5020763} {\bibfield  {journal} {\bibinfo  {journal} {AIP
  Advances}\ }\textbf {\bibinfo {volume} {8}},\ \bibinfo {pages} {025318}
  (\bibinfo {year} {2018})}\BibitemShut {NoStop}%
\bibitem [{\citenamefont {Moll}\ \emph {et~al.}(2016)\citenamefont {Moll},
  \citenamefont {Kushwaha}, \citenamefont {Nandi}, \citenamefont {Schmidt},\
  and\ \citenamefont {Mackenzie}}]{moll_evidence_2016}%
  \BibitemOpen
  \bibfield  {author} {\bibinfo {author} {\bibfnamefont {P.~J.~W.}\
  \bibnamefont {Moll}}, \bibinfo {author} {\bibfnamefont {P.}~\bibnamefont
  {Kushwaha}}, \bibinfo {author} {\bibfnamefont {N.}~\bibnamefont {Nandi}},
  \bibinfo {author} {\bibfnamefont {B.}~\bibnamefont {Schmidt}}, \ and\
  \bibinfo {author} {\bibfnamefont {A.~P.}\ \bibnamefont {Mackenzie}},\ }\href
  {\doibase 10.1126/science.aac8385} {\bibfield  {journal} {\bibinfo  {journal}
  {Science}\ }\textbf {\bibinfo {volume} {351}},\ \bibinfo {pages} {1061}
  (\bibinfo {year} {2016})}\BibitemShut {NoStop}%
\bibitem [{\citenamefont {Gooth}\ \emph {et~al.}(2017)\citenamefont {Gooth},
  \citenamefont {Menges}, \citenamefont {Shekhar}, \citenamefont {S{\"u}{\ss}},
  \citenamefont {Kumar}, \citenamefont {Sun}, \citenamefont {Drechsler},
  \citenamefont {Zierold}, \citenamefont {Felser},\ and\ \citenamefont
  {Gotsmann}}]{gooth_electrical_2017}%
  \BibitemOpen
  \bibfield  {author} {\bibinfo {author} {\bibfnamefont {J.}~\bibnamefont
  {Gooth}}, \bibinfo {author} {\bibfnamefont {F.}~\bibnamefont {Menges}},
  \bibinfo {author} {\bibfnamefont {C.}~\bibnamefont {Shekhar}}, \bibinfo
  {author} {\bibfnamefont {V.}~\bibnamefont {S{\"u}{\ss}}}, \bibinfo {author}
  {\bibfnamefont {N.}~\bibnamefont {Kumar}}, \bibinfo {author} {\bibfnamefont
  {Y.}~\bibnamefont {Sun}}, \bibinfo {author} {\bibfnamefont {U.}~\bibnamefont
  {Drechsler}}, \bibinfo {author} {\bibfnamefont {R.}~\bibnamefont {Zierold}},
  \bibinfo {author} {\bibfnamefont {C.}~\bibnamefont {Felser}}, \ and\ \bibinfo
  {author} {\bibfnamefont {B.}~\bibnamefont {Gotsmann}},\ }\href
  {http://arxiv.org/abs/1706.05925} {\bibfield  {journal} {\bibinfo  {journal}
  {arXiv:1706.05925 [cond-mat]}\ } (\bibinfo {year} {2017})},\ \bibinfo {note}
  {arXiv: 1706.05925}\BibitemShut {NoStop}%
\bibitem [{\citenamefont {Torre}\ \emph {et~al.}(2015)\citenamefont {Torre},
  \citenamefont {Tomadin}, \citenamefont {Geim},\ and\ \citenamefont
  {Polini}}]{torre_nonlocal_2015}%
  \BibitemOpen
  \bibfield  {author} {\bibinfo {author} {\bibfnamefont {I.}~\bibnamefont
  {Torre}}, \bibinfo {author} {\bibfnamefont {A.}~\bibnamefont {Tomadin}},
  \bibinfo {author} {\bibfnamefont {A.~K.}\ \bibnamefont {Geim}}, \ and\
  \bibinfo {author} {\bibfnamefont {M.}~\bibnamefont {Polini}},\ }\href
  {\doibase 10.1103/PhysRevB.92.165433} {\bibfield  {journal} {\bibinfo
  {journal} {Phys. Rev. B}\ }\textbf {\bibinfo {volume} {92}},\ \bibinfo
  {pages} {165433} (\bibinfo {year} {2015})}\BibitemShut {NoStop}%
\bibitem [{\citenamefont {Levitov}\ and\ \citenamefont
  {Falkovich}(2016)}]{levitov_electron_2016}%
  \BibitemOpen
  \bibfield  {author} {\bibinfo {author} {\bibfnamefont {L.}~\bibnamefont
  {Levitov}}\ and\ \bibinfo {author} {\bibfnamefont {G.}~\bibnamefont
  {Falkovich}},\ }\href {\doibase 10.1038/nphys3667} {\bibfield  {journal}
  {\bibinfo  {journal} {Nature Physics}\ }\textbf {\bibinfo {volume} {12}},\
  \bibinfo {pages} {672} (\bibinfo {year} {2016})}\BibitemShut {NoStop}%
\bibitem [{\citenamefont {Pellegrino}\ \emph {et~al.}(2016)\citenamefont
  {Pellegrino}, \citenamefont {Torre}, \citenamefont {Geim},\ and\
  \citenamefont {Polini}}]{pellegrino_electron_2016}%
  \BibitemOpen
  \bibfield  {author} {\bibinfo {author} {\bibfnamefont {F.~M.~D.}\
  \bibnamefont {Pellegrino}}, \bibinfo {author} {\bibfnamefont
  {I.}~\bibnamefont {Torre}}, \bibinfo {author} {\bibfnamefont {A.~K.}\
  \bibnamefont {Geim}}, \ and\ \bibinfo {author} {\bibfnamefont
  {M.}~\bibnamefont {Polini}},\ }\href {\doibase 10.1103/PhysRevB.94.155414}
  {\bibfield  {journal} {\bibinfo  {journal} {Phys. Rev. B}\ }\textbf {\bibinfo
  {volume} {94}},\ \bibinfo {pages} {155414} (\bibinfo {year}
  {2016})}\BibitemShut {NoStop}%
\bibitem [{\citenamefont {Bandurin}\ \emph {et~al.}(2018)\citenamefont
  {Bandurin}, \citenamefont {Shytov}, \citenamefont {Falkovich}, \citenamefont
  {Kumar}, \citenamefont {Shalom}, \citenamefont {Grigorieva}, \citenamefont
  {Geim},\ and\ \citenamefont {Levitov}}]{bandurin_probing_2018}%
  \BibitemOpen
  \bibfield  {author} {\bibinfo {author} {\bibfnamefont {D.~A.}\ \bibnamefont
  {Bandurin}}, \bibinfo {author} {\bibfnamefont {A.~V.}\ \bibnamefont
  {Shytov}}, \bibinfo {author} {\bibfnamefont {G.}~\bibnamefont {Falkovich}},
  \bibinfo {author} {\bibfnamefont {R.~K.}\ \bibnamefont {Kumar}}, \bibinfo
  {author} {\bibfnamefont {M.~B.}\ \bibnamefont {Shalom}}, \bibinfo {author}
  {\bibfnamefont {I.~V.}\ \bibnamefont {Grigorieva}}, \bibinfo {author}
  {\bibfnamefont {A.~K.}\ \bibnamefont {Geim}}, \ and\ \bibinfo {author}
  {\bibfnamefont {L.~S.}\ \bibnamefont {Levitov}},\ }\href
  {http://arxiv.org/abs/1806.03231} {\bibfield  {journal} {\bibinfo  {journal}
  {arXiv:1806.03231 [cond-mat]}\ } (\bibinfo {year} {2018})}\BibitemShut
  {NoStop}%
\bibitem [{\citenamefont {Shytov}\ \emph {et~al.}(2018)\citenamefont {Shytov},
  \citenamefont {Kong}, \citenamefont {Falkovich},\ and\ \citenamefont
  {Levitov}}]{shytov_electron_2018}%
  \BibitemOpen
  \bibfield  {author} {\bibinfo {author} {\bibfnamefont {A.}~\bibnamefont
  {Shytov}}, \bibinfo {author} {\bibfnamefont {J.~F.}\ \bibnamefont {Kong}},
  \bibinfo {author} {\bibfnamefont {G.}~\bibnamefont {Falkovich}}, \ and\
  \bibinfo {author} {\bibfnamefont {L.}~\bibnamefont {Levitov}},\ }\href
  {http://arxiv.org/abs/1806.09538} {\bibfield  {journal} {\bibinfo  {journal}
  {arXiv:1806.09538 [cond-mat]}\ } (\bibinfo {year} {2018})}\BibitemShut
  {NoStop}%
\bibitem [{\citenamefont {Eriksson}\ \emph
  {et~al.}(1996{\natexlab{a}})\citenamefont {Eriksson}, \citenamefont {Beck},
  \citenamefont {Topinka}, \citenamefont {Katine}, \citenamefont {Westervelt},
  \citenamefont {Campman},\ and\ \citenamefont
  {Gossard}}]{eriksson_cryogenic_1996}%
  \BibitemOpen
  \bibfield  {author} {\bibinfo {author} {\bibfnamefont {M.~A.}\ \bibnamefont
  {Eriksson}}, \bibinfo {author} {\bibfnamefont {R.~G.}\ \bibnamefont {Beck}},
  \bibinfo {author} {\bibfnamefont {M.}~\bibnamefont {Topinka}}, \bibinfo
  {author} {\bibfnamefont {J.~A.}\ \bibnamefont {Katine}}, \bibinfo {author}
  {\bibfnamefont {R.~M.}\ \bibnamefont {Westervelt}}, \bibinfo {author}
  {\bibfnamefont {K.~L.}\ \bibnamefont {Campman}}, \ and\ \bibinfo {author}
  {\bibfnamefont {A.~C.}\ \bibnamefont {Gossard}},\ }\href {\doibase
  10.1063/1.117801} {\bibfield  {journal} {\bibinfo  {journal} {Applied Physics
  Letters}\ }\textbf {\bibinfo {volume} {69}},\ \bibinfo {pages} {671}
  (\bibinfo {year} {1996}{\natexlab{a}})}\BibitemShut {NoStop}%
\bibitem [{\citenamefont {Topinka}\ \emph {et~al.}(2000)\citenamefont
  {Topinka}, \citenamefont {LeRoy}, \citenamefont {Shaw}, \citenamefont
  {Heller}, \citenamefont {Westervelt}, \citenamefont {Maranowski},\ and\
  \citenamefont {Gossard}}]{topinka_imaging_2000}%
  \BibitemOpen
  \bibfield  {author} {\bibinfo {author} {\bibfnamefont {M.~A.}\ \bibnamefont
  {Topinka}}, \bibinfo {author} {\bibfnamefont {B.~J.}\ \bibnamefont {LeRoy}},
  \bibinfo {author} {\bibfnamefont {S.~E.~J.}\ \bibnamefont {Shaw}}, \bibinfo
  {author} {\bibfnamefont {E.~J.}\ \bibnamefont {Heller}}, \bibinfo {author}
  {\bibfnamefont {R.~M.}\ \bibnamefont {Westervelt}}, \bibinfo {author}
  {\bibfnamefont {K.~D.}\ \bibnamefont {Maranowski}}, \ and\ \bibinfo {author}
  {\bibfnamefont {A.~C.}\ \bibnamefont {Gossard}},\ }\href {\doibase
  10.1126/science.289.5488.2323} {\bibfield  {journal} {\bibinfo  {journal}
  {Science}\ }\textbf {\bibinfo {volume} {289}},\ \bibinfo {pages} {2323}
  (\bibinfo {year} {2000})}\BibitemShut {NoStop}%
\bibitem [{\citenamefont {Berl}\ \emph {et~al.}(2016)\citenamefont {Berl},
  \citenamefont {Tiemann}, \citenamefont {Dietsche}, \citenamefont {Karl},\
  and\ \citenamefont {Wegscheider}}]{berl_structured_2016}%
  \BibitemOpen
  \bibfield  {author} {\bibinfo {author} {\bibfnamefont {M.}~\bibnamefont
  {Berl}}, \bibinfo {author} {\bibfnamefont {L.}~\bibnamefont {Tiemann}},
  \bibinfo {author} {\bibfnamefont {W.}~\bibnamefont {Dietsche}}, \bibinfo
  {author} {\bibfnamefont {H.}~\bibnamefont {Karl}}, \ and\ \bibinfo {author}
  {\bibfnamefont {W.}~\bibnamefont {Wegscheider}},\ }\href {\doibase
  10.1063/1.4945090} {\bibfield  {journal} {\bibinfo  {journal} {Applied
  Physics Letters}\ }\textbf {\bibinfo {volume} {108}},\ \bibinfo {pages}
  {132102} (\bibinfo {year} {2016})}\BibitemShut {NoStop}%
\bibitem [{\citenamefont {Jungwirth}\ and\ \citenamefont
  {MacDonald}(1996)}]{jungwirth_electron-electron_1996}%
  \BibitemOpen
  \bibfield  {author} {\bibinfo {author} {\bibfnamefont {T.}~\bibnamefont
  {Jungwirth}}\ and\ \bibinfo {author} {\bibfnamefont {A.~H.}\ \bibnamefont
  {MacDonald}},\ }\href {\doibase 10.1103/PhysRevB.53.7403} {\bibfield
  {journal} {\bibinfo  {journal} {Phys. Rev. B}\ }\textbf {\bibinfo {volume}
  {53}},\ \bibinfo {pages} {7403} (\bibinfo {year} {1996})}\BibitemShut
  {NoStop}%
\bibitem [{\citenamefont {Giuliani}\ and\ \citenamefont
  {Vignale}(2008)}]{giuliani_quantum_2008}%
  \BibitemOpen
  \bibfield  {author} {\bibinfo {author} {\bibfnamefont {G.}~\bibnamefont
  {Giuliani}}\ and\ \bibinfo {author} {\bibfnamefont {G.}~\bibnamefont
  {Vignale}},\ }\href@noop {} {\emph {\bibinfo {title} {Quantum {Theory} of the
  {Electron} {Liquid}}}},\ \bibinfo {edition} {1st}\ ed.\ (\bibinfo
  {publisher} {Cambridge University Press},\ \bibinfo {address} {Cambridge},\
  \bibinfo {year} {2008})\BibitemShut {NoStop}%
\bibitem [{\citenamefont {Crook}\ \emph {et~al.}(2000)\citenamefont {Crook},
  \citenamefont {Smith}, \citenamefont {Simmons},\ and\ \citenamefont
  {Ritchie}}]{crook_imaging_2000}%
  \BibitemOpen
  \bibfield  {author} {\bibinfo {author} {\bibfnamefont {R.}~\bibnamefont
  {Crook}}, \bibinfo {author} {\bibfnamefont {C.~G.}\ \bibnamefont {Smith}},
  \bibinfo {author} {\bibfnamefont {M.~Y.}\ \bibnamefont {Simmons}}, \ and\
  \bibinfo {author} {\bibfnamefont {D.~A.}\ \bibnamefont {Ritchie}},\ }\href
  {\doibase 10.1103/PhysRevB.62.5174} {\bibfield  {journal} {\bibinfo
  {journal} {Phys. Rev. B}\ }\textbf {\bibinfo {volume} {62}},\ \bibinfo
  {pages} {5174} (\bibinfo {year} {2000})}\BibitemShut {NoStop}%
\bibitem [{\citenamefont {Aidala}\ \emph {et~al.}(2007)\citenamefont {Aidala},
  \citenamefont {Parrott}, \citenamefont {Kramer}, \citenamefont {Heller},
  \citenamefont {Westervelt}, \citenamefont {Hanson},\ and\ \citenamefont
  {Gossard}}]{aidala_imaging_2007}%
  \BibitemOpen
  \bibfield  {author} {\bibinfo {author} {\bibfnamefont {K.~E.}\ \bibnamefont
  {Aidala}}, \bibinfo {author} {\bibfnamefont {R.~E.}\ \bibnamefont {Parrott}},
  \bibinfo {author} {\bibfnamefont {T.}~\bibnamefont {Kramer}}, \bibinfo
  {author} {\bibfnamefont {E.~J.}\ \bibnamefont {Heller}}, \bibinfo {author}
  {\bibfnamefont {R.~M.}\ \bibnamefont {Westervelt}}, \bibinfo {author}
  {\bibfnamefont {M.~P.}\ \bibnamefont {Hanson}}, \ and\ \bibinfo {author}
  {\bibfnamefont {A.~C.}\ \bibnamefont {Gossard}},\ }\href {\doibase
  10.1038/nphys628} {\bibfield  {journal} {\bibinfo  {journal} {Nature
  Physics}\ }\textbf {\bibinfo {volume} {3}},\ \bibinfo {pages} {464} (\bibinfo
  {year} {2007})}\BibitemShut {NoStop}%
\bibitem [{\citenamefont {Bhandari}\ \emph {et~al.}(2016)\citenamefont
  {Bhandari}, \citenamefont {Lee}, \citenamefont {Klales}, \citenamefont
  {Watanabe}, \citenamefont {Taniguchi}, \citenamefont {Heller}, \citenamefont
  {Kim},\ and\ \citenamefont {Westervelt}}]{bhandari_imaging_2016}%
  \BibitemOpen
  \bibfield  {author} {\bibinfo {author} {\bibfnamefont {S.}~\bibnamefont
  {Bhandari}}, \bibinfo {author} {\bibfnamefont {G.-H.}\ \bibnamefont {Lee}},
  \bibinfo {author} {\bibfnamefont {A.}~\bibnamefont {Klales}}, \bibinfo
  {author} {\bibfnamefont {K.}~\bibnamefont {Watanabe}}, \bibinfo {author}
  {\bibfnamefont {T.}~\bibnamefont {Taniguchi}}, \bibinfo {author}
  {\bibfnamefont {E.}~\bibnamefont {Heller}}, \bibinfo {author} {\bibfnamefont
  {P.}~\bibnamefont {Kim}}, \ and\ \bibinfo {author} {\bibfnamefont {R.~M.}\
  \bibnamefont {Westervelt}},\ }\href {\doibase 10.1021/acs.nanolett.5b04609}
  {\bibfield  {journal} {\bibinfo  {journal} {Nano Lett.}\ }\textbf {\bibinfo
  {volume} {16}},\ \bibinfo {pages} {1690} (\bibinfo {year}
  {2016})}\BibitemShut {NoStop}%
\bibitem [{\citenamefont {Eriksson}\ \emph
  {et~al.}(1996{\natexlab{b}})\citenamefont {Eriksson}, \citenamefont {Beck},
  \citenamefont {Topinka}, \citenamefont {Katine}, \citenamefont {Westervelt},
  \citenamefont {Campman},\ and\ \citenamefont
  {Gossard}}]{eriksson_effect_1996}%
  \BibitemOpen
  \bibfield  {author} {\bibinfo {author} {\bibfnamefont {M.~A.}\ \bibnamefont
  {Eriksson}}, \bibinfo {author} {\bibfnamefont {R.~G.}\ \bibnamefont {Beck}},
  \bibinfo {author} {\bibfnamefont {M.~A.}\ \bibnamefont {Topinka}}, \bibinfo
  {author} {\bibfnamefont {J.~A.}\ \bibnamefont {Katine}}, \bibinfo {author}
  {\bibfnamefont {R.~M.}\ \bibnamefont {Westervelt}}, \bibinfo {author}
  {\bibfnamefont {K.~L.}\ \bibnamefont {Campman}}, \ and\ \bibinfo {author}
  {\bibfnamefont {A.~C.}\ \bibnamefont {Gossard}},\ }\href {\doibase
  10.1006/spmi.1996.0100} {\bibfield  {journal} {\bibinfo  {journal}
  {Superlattices and Microstructures}\ }\textbf {\bibinfo {volume} {20}},\
  \bibinfo {pages} {435} (\bibinfo {year} {1996}{\natexlab{b}})}\BibitemShut
  {NoStop}%
\bibitem [{\citenamefont {Topinka}\ \emph {et~al.}(2001)\citenamefont
  {Topinka}, \citenamefont {LeRoy}, \citenamefont {Westervelt}, \citenamefont
  {Shaw}, \citenamefont {Fleischmann}, \citenamefont {Heller}, \citenamefont
  {Maranowski},\ and\ \citenamefont {Gossard}}]{topinka_coherent_2001}%
  \BibitemOpen
  \bibfield  {author} {\bibinfo {author} {\bibfnamefont {M.~A.}\ \bibnamefont
  {Topinka}}, \bibinfo {author} {\bibfnamefont {B.~J.}\ \bibnamefont {LeRoy}},
  \bibinfo {author} {\bibfnamefont {R.~M.}\ \bibnamefont {Westervelt}},
  \bibinfo {author} {\bibfnamefont {S.~E.~J.}\ \bibnamefont {Shaw}}, \bibinfo
  {author} {\bibfnamefont {R.}~\bibnamefont {Fleischmann}}, \bibinfo {author}
  {\bibfnamefont {E.~J.}\ \bibnamefont {Heller}}, \bibinfo {author}
  {\bibfnamefont {K.~D.}\ \bibnamefont {Maranowski}}, \ and\ \bibinfo {author}
  {\bibfnamefont {A.~C.}\ \bibnamefont {Gossard}},\ }\href {\doibase
  10.1038/35065553} {\bibfield  {journal} {\bibinfo  {journal} {Nature}\
  }\textbf {\bibinfo {volume} {410}},\ \bibinfo {pages} {183} (\bibinfo {year}
  {2001})}\BibitemShut {NoStop}%
\bibitem [{\citenamefont {Beenakker}\ and\ \citenamefont {van
  Houten}(1991)}]{beenakker_quantum_1991}%
  \BibitemOpen
  \bibfield  {author} {\bibinfo {author} {\bibfnamefont {C.~W.~J.}\
  \bibnamefont {Beenakker}}\ and\ \bibinfo {author} {\bibfnamefont
  {H.}~\bibnamefont {van Houten}},\ }\href {\doibase
  10.1016/S0081-1947(08)60091-0} {\bibfield  {journal} {\bibinfo  {journal}
  {Solid State Physics}\ }\textbf {\bibinfo {volume} {44}},\ \bibinfo {pages}
  {1} (\bibinfo {year} {1991})}\BibitemShut {NoStop}%
\bibitem [{\citenamefont {Giuliani}\ and\ \citenamefont
  {Quinn}(1982)}]{giuliani_lifetime_1982}%
  \BibitemOpen
  \bibfield  {author} {\bibinfo {author} {\bibfnamefont {G.~F.}\ \bibnamefont
  {Giuliani}}\ and\ \bibinfo {author} {\bibfnamefont {J.~J.}\ \bibnamefont
  {Quinn}},\ }\href {\doibase 10.1103/PhysRevB.26.4421} {\bibfield  {journal}
  {\bibinfo  {journal} {Phys. Rev. B}\ }\textbf {\bibinfo {volume} {26}},\
  \bibinfo {pages} {4421} (\bibinfo {year} {1982})}\BibitemShut {NoStop}%
\end{thebibliography}%

\pagebreak
\widetext
\begin{center}
\textbf{\large Supplemental Materials: Scanning Gate Microscopy in a \\ Viscous Electron Fluid}
\end{center}
\setcounter{equation}{0}
\setcounter{figure}{0}
\setcounter{table}{0}
\setcounter{page}{1}
\makeatletter
\renewcommand{\theequation}{S\arabic{equation}}
\renewcommand{\thefigure}{S\arabic{figure}}
\renewcommand*{\thepage}{S\arabic{page}}
\renewcommand{\bibnumfmt}[1]{[S#1]}
\renewcommand{\citenumfont}[1]{S#1}

This supplemental material contains information exceeding the scope of the main text. We provide details about the experimental methods and results of the electron transport in the bulk 2DEG. Furthermore, we show the absolute values of the electron-electron scattering length and the Drude scattering length as a function of the parameters $T$ and $n$. The last two sections describe details of the hydrodynamic model and the classical trajectory calculations.

%

\tableofcontents

\section{Experimental methods}
We use a Ga[Al]As heterostructure with a 2DEG buried 130~nm below the surface and a grown back-gate \SI{1.13}{\micro\meter} below the 2DEG \cite{berl_structured_2016}. A voltage $V_\mathrm{bg}$ applied to the back-gate tunes the bulk electron density according to $n=(1.21 + V_\mathrm{bg}/\SI{1.67}{V} )\times 10^{11} \, \si{\per\square\centi\meter}$ (see the third section for details). At $V_\mathrm{bg}=\SI{0}{V}$ the electron mobility is \SI{6.2e6}{cm^2/Vs} at \SI{32}{\milli\kelvin}.
On this heterostructure we define \SI{35}{nm} high TiAu top-gates by electron-beam lithography. To deplete the 2DEG underneath, we apply a gate voltage of $V_\mathrm{top-gates}=-0.18 \times V_\mathrm{bg} - \SI{0.5}{V}$ with respect to the 2DEG potential ($n=\SI{1.2e12}{cm^{-2}}$: $V_\mathrm{bg} = \SI{0.0}{V}$ and $V_\mathrm{tip}= - \SI{8.0}{V}$, $n=\SI{0.3e12}{cm^{-2}}$: $V_\mathrm{bg} = - \SI{1.5}{V}$ and $V_\mathrm{tip}= - \SI{2.75}{V}$) . 

\begin{figure}
 \includegraphics[width=0.8\linewidth]{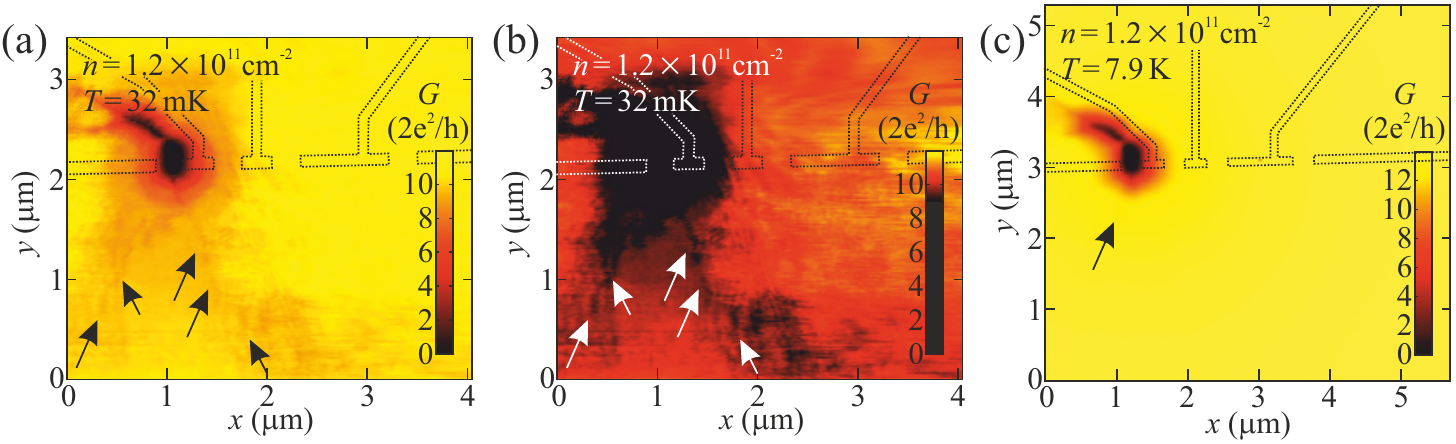}
 \caption{Two-terminal conductance $G$ of the source orifice measured while measuring the vicinity resistances presented in Figs.~2(b) and~(c) in the main text. (a),(b) $G$ measured at base temperature with two different color scale settings to highlight the weak pattern of branched electron flow marked by the arrows. (c) $G$ at $T=\SI{7.9}{K}$ showing similar behavior as in (a) and thus confirming the same invasiveness of the tip-induced potential as at base temperature. Due to the high temperature the branch pattern is reduced to a weak dip marked by the arrow.
 }
 \label{fig_branches}
\end{figure}

To keep the tip-induced potential roughly proportional to the Fermi energy at all electron densities, we apply a tip voltage $V_\mathrm{tip}=\SI{-8}{\volt}-3.5 \times V_\mathrm{bg}$ with respect to the 2DEG potential. Such a negative $V_\mathrm{tip}$ depletes the 2DEG below the tip, which is supported by the observation of the pattern of branched electron flow \cite{topinka_coherent_2001} at base temperature. The corresponding maps of the two-terminal conductance as a function of tip position are shown in Fig.~\ref{fig_branches}. 

\begin{figure}
\begin{center}
 \includegraphics[width=0.35\linewidth]{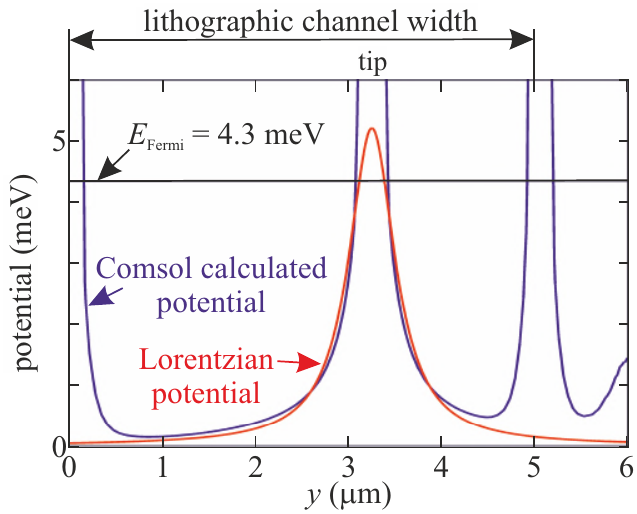}
\end{center}
\caption{The blue line shows the Comsol-calculated potential from Fig.~5(a) of the main text (vertical cut through the tip position). Red shows the Lorentzian approximation of the potential as described by eq.~\eqref{eq_Lorentz}. The Fermi energy is marked by the black horizontal line.}
 \label{label_comsolcut}
\end{figure}

We estimate the tip depletion diameter to be approximately \SI{300}{nm} from choosing $V_\mathrm{tip}$ more negative than the depleting voltage. The finite element simulations used for the ballistic model (Fig.~5 in the main text) confirm this estimate: Figure~\ref{label_comsolcut} shows a vertical cut through the tip position and we find that the electrostatic potential (blue) exceeds the Fermi energy over a distance of approximately \SI{300}{nm}. Outside the depletion disk, the tip induced potential approaches zero within a distance of \SI{1}{\micro\meter} around the tip. Because the local electron density in the 2DEG is proportional to the difference of the Fermi energy and the electrostatic potential, it is reduced with respect to the bulk value within \SI{1}{\micro\meter} around the tip. At larger distances, the tip induced density modulation is smaller than fluctuations expected from the random background potential present in GaAs 2DEGs. 
The red line in Fig.~\ref{label_comsolcut} indicates the Lorentzian potential describing the tip in the hydrodynamic model described in the last section.

Each of the measurement cables to the source and drain contact (see Fig.~1(a) of the main text) has a resistance of \SI{10}{\kilo\ohm} from the cold RC-filter. To determine the two-terminal resistance of the current injector orifice, we use two additional measurement leads which allow for the current-free measurement of the voltage between source and drain contact. 

To remain in the linear transport regime, we apply a small voltage of \SI{100}{\micro\volt} to the room temperature ends of the cables. In the absence of the tip, the cable resistance ($2\times \SI{10}{\kilo\ohm}$) dominates over the two-terminal resistance of the injector (depending on the electron density: \SI{1.5}{\kilo\ohm} to \SI{10}{\kilo\ohm}). Therefore the voltage between source and drain contact is much smaller than the applied \SI{100}{\micro\volt} and the current $I$ is limited by the filter resistors to $I_\mathrm{max}=\SI{100}{\micro\volt}/(2\times \SI{10}{\kilo\ohm})=\SI{5}{nA}$. When the tip approaches and depletes the current injector, we reach $I=0$ and the voltage between source and drain contact is \SI{100}{\micro\volt}. 
At base temperature, we observe at least 6 conductance plateaus conductance of each of the four orifices from the modes of the quantum point contacts. At $T>\SI{4}{K}$ the quantum point contact modes are obscured by thermal smearing.

\newpage
\section{Controlling the sample temperature}
In experiments at low temperatures, differences between the electronic temperature of the sample and the cryostat can arise. In the following we will describe in detail, how we determined the temperature, how the cryostat temperature was controlled, and why we can assume to have only a negligible difference between the electronic sample temperature and the temperature measured on the sample stage. 

\begin{figure}
\begin{center}
 \includegraphics[width=0.7\linewidth]{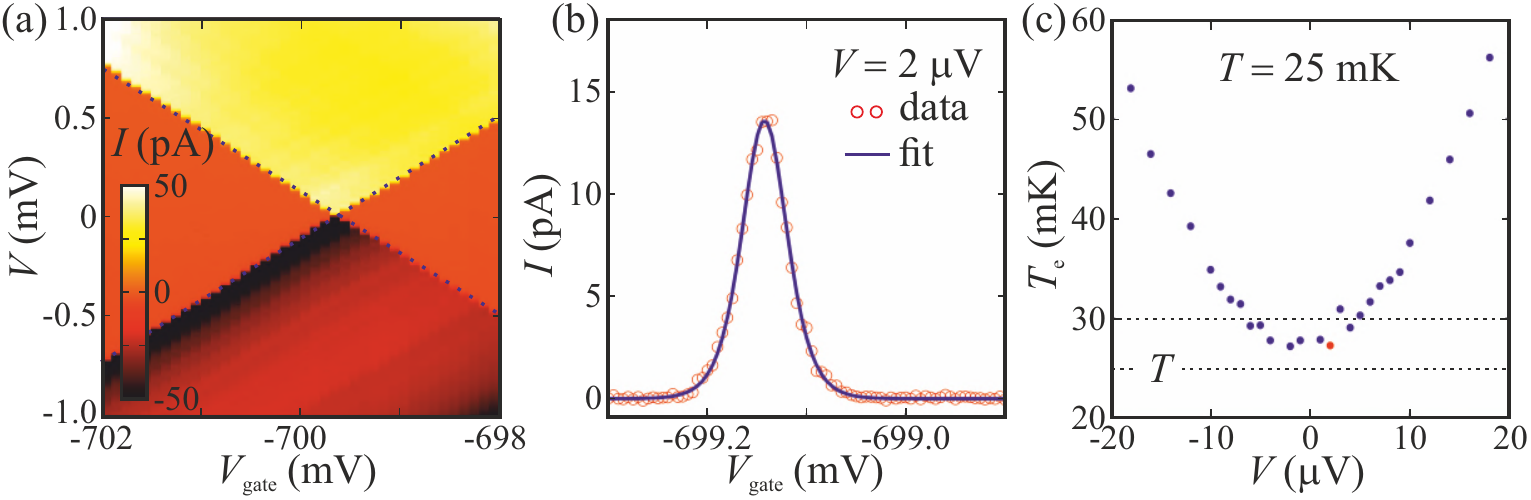}
\end{center}
\caption{Coulomb blockade measurement of a quantum dot in the SGM setup to determine the electron temperature at a cryostat temperature of $T=\SI{25}{K}$. (a) Sample current $I$ as a function of the plunger gate voltage $V_\mathrm{gate}$ and the bias voltage $V$ between source and drain. (b) High-resolution measurement of $I$ as a function of $V_\mathrm{gate}$ (red circles) at $V=\SI{2}{\micro\volt}$. The blue line is the fit according to eq.~\eqref{eq_coulombpeak} to extract the electron temperature. (c) $T_\mathrm{e}$ from fitting $I(V_\mathrm{gate})$ as a function of $V$ lie below \SI{30}{mK} at small source-drain voltage. The red dot marks the example in panel (b).
}
 \label{label_coulomb}
\end{figure}
We characterized the electronic temperature in our SGM setup by measuring Coulomb blockade resonances (using a different GaAs sample with a top-gate defined quantum dot)  with exactly the same wiring and filtering of the electric signals. Figure~\ref{label_coulomb} describes how we extract the electron temperature $T_\mathrm{e}$ from fitting the current $I$ across a Coulomb blockade resonance with
\begin{equation}
I(V_\mathrm{gate})=\frac{I_0}{\mathrm{cosh}^2(\alpha_\mathrm{gate}(V_\mathrm{gate}-V_0)/2k_\mathrm{B}T_\mathrm{e})} 
\label{eq_coulombpeak}
\end{equation}
with $I_0$ and $V_0$ being the current and the gate voltage at the maximum of $I(V_\mathrm{gate})$ following Ref.~\citenum{beenakker_quantum_1991}. The lever arm $\alpha_\mathrm{gate}$ describes the capacity of the gate to the quantum dot, its value is extracted from the slope of the blue dotted lines in Fig.~\ref{label_coulomb}(a). From fitting as shown in Fig.~\ref{label_coulomb}(b) we obtain the electron temperature as a function of source-drain voltage $V$ as shown in Fig.~\ref{label_coulomb}(c).
At a mixing chamber temperature of \SI{25}{mK} (according to the Oxford Instruments thermometry), we extract an electronic temperature below \SI{30}{mK} thanks to our improvements of the thermal anchoring of the cabling. The AFM cabling is thermally anchored similarly to the sample cabling and there is no indication of additional heating due to the presence of the tip. This supports our assumption to have only a negligible difference between electronic temperature and the cryostat temperature at millikelvin temperatures. At higher temperatures, the difference of electronic temperature and cryostat temperature typically decreases due to the better thermal conductance across material interfaces and through insulating materials (which are the main two problems for thermal anchoring at millikelvin temperatures).

For the measurements at base temperature, we rely on the small difference between electron temperature and cryostat temperature shown in the Coulomb blockade measurements. In the following we describe our measures to reach higher temperatures, e.g. \SI{7.9}{K} for the measurements in Figs.~2(c),(d) and Fig.~3 of the main text. First, we withdraw the mixture from the dilution unit. Second, we determine the sample temperature by measuring the resistance of a Lakeshore RX-202A RuOx thermometer mounted to the sample stage by standard lock-in technique. This thermometer is separated by less than \SI{1}{cm} from the chip carrier and by approximately \SI{28}{cm} from the mixing chamber plate. Third, we heat the mixing chamber plate using the heater installed by Oxford instruments, but a software-controllable voltage source. A PI-controller controls the heater voltage to achieve the desired   sample temperature. We achieve a temperature stability of $\pm \SI{5}{mK}$ at \SI{7.9}{K}, which is limited by the resolution of the lock-in amplifier measuring the RuOx resistor. To obtain the measurements at \SI{7.9}{K}, we first heat the cryostat to this temperature for two days to ensure thermalization of the AFM components. Then we approach the tip to the sample and wait for additional \SI{36}{h} before starting the measurement to avoid drifts and tip crashes. This slow procedure ensures a small temperature difference between the measured temperature and the electronic temperature of the sample. 

For the measurement as a function of temperature shown in Fig.~1(b), we do not control the temperature to each single point but heat the mixing chamber plate slowly and record the sample stage temperature together with the measurement data. The measurement shown in Fig.~1(b) has been obtained during a slow warm-up from \SI{32}{mK} to \SI{30}{K} over \SI{24}{h} to ensure a slow heating and good thermalization of the sample and the sample stage.

\section{Electron density and mobility as a function of back-gate voltage at base temperature}
\begin{figure}
 \includegraphics[width=0.45\linewidth]{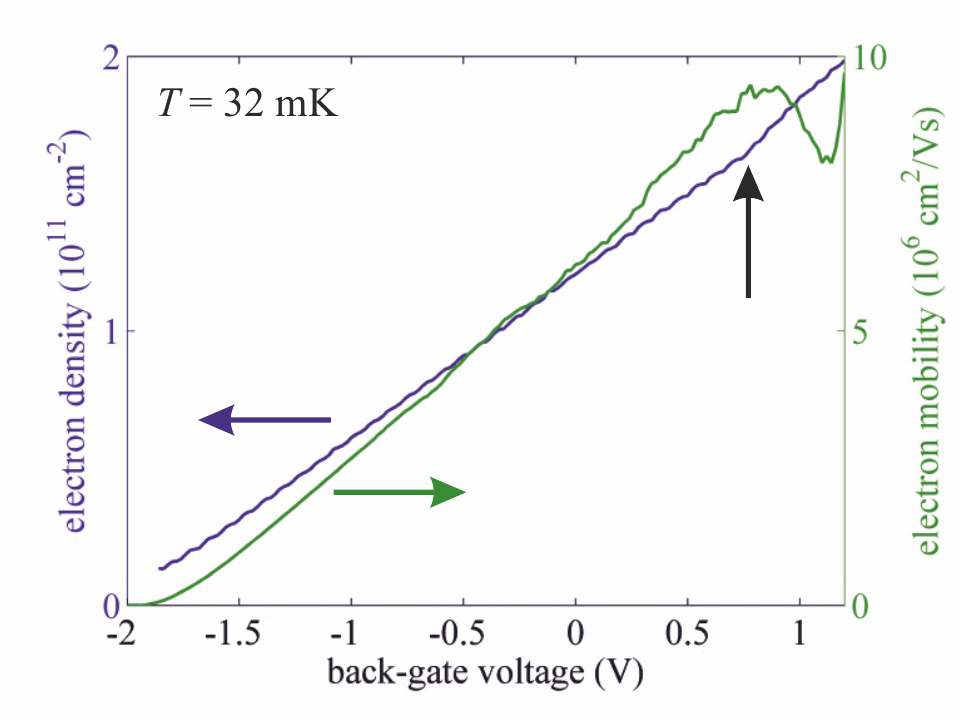}
 \caption{Hall density and electron mobility as a function of $V_\mathrm{bg}$ at $T=\SI{32}{\milli\kelvin}$. The vertical arrow marks the onset of a second 2DEG forming at $V_\mathrm{bg}>\SI{0.8}{V}$. 
 }
 \label{fig_mobi_dens}
\end{figure}

We extract the electron density and mobility in Fig.~\ref{fig_mobi_dens} from standard longitudinal and Hall resistance measurements at base temperature. For back-gate voltages in the range $V_\mathrm{bg}=\SI{-1.8}{V}$ to \SI{0.8}{V} we find the linear increase of $n(V_\mathrm{bg})$ as expected from the parallel-plate capacitor model and a monotonic increase of the mobility. At $V_\mathrm{bg}<\SI{-1.8}{V}$ the electrons localize and the 2DEG is insulating. At high $V_\mathrm{bg}>\SI{0.8}{V}$ a second 2DEG layer forms at a heterostructure interface between 2DEG and back-gate and we observe a decrease of the mobility. 
The measurements in the main text are obtained at $V_\mathrm{bg}=\SI{0}{\volt}$ ($n=\SI{1.2e11}{cm^{-2}}$) and at $V_\mathrm{bg}=\SI{-1.5}{\volt}$ ($n=\SI{0.3e11}{cm^{-2}}$). 

\newpage
\section{Drude mean free path and electron-electron scattering length as a function of temperature and electron density}
To observe viscous effects, the momentum relaxation length $l_{\mathrm{D}}$ must exceed the electron-electron scattering length $l_{\mathrm{ee}}$. This section presents the experimental values of $l_{\mathrm{D}}$ and calculated values for $l_{\mathrm{ee}}$.

\begin{figure}
 \includegraphics[width=0.55\linewidth]{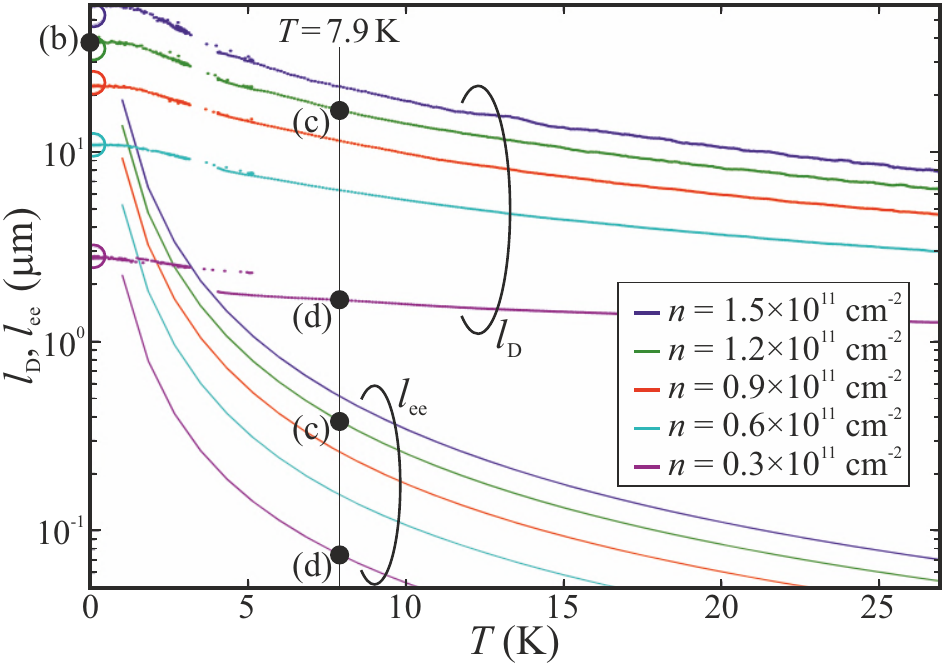}
 \caption{
 Drude scattering length $l_{\mathrm{D}}$ extracted from three measurements covering different temperature ranges, namely, at 30~mK, 30~mK~-~4.2~K and $4.2-27\, \mathrm{K}$ together with the numerically calculated electron-electron interaction length \cite{jungwirth_electron-electron_1996}. The presented SGM data in the main text is measured at electron densities of $\SI{1.2e11}{cm^{-2}}$ and $\SI{0.3e11}{cm^{-2}}$. For all densities, $l_{\mathrm{D}}$ and $l_\mathrm{ee}$ are comparable at 1.5~K. At higher temperatures $l_{\mathrm{D}}$ exceeds $l_\mathrm{ee}$ and viscous effects can arise. The black dots mark the parameters of the SGM images in Fig.~2 of the main text. 
}
 \label{figS2label}
\end{figure}
Figure~\ref{figS2label} shows the measured Drude scattering length $l_\mathrm{D}$ and the calculated electron-electron interaction length $l_\mathrm{ee}$. $l_\mathrm{D}$ is extracted from longitudinal and Hall resistance measurements, analogously to the electron density and the electron mobility in Fig.~\ref{fig_mobi_dens}. We calculate the electron-electron scattering length $l_\mathrm{ee}=v_\mathrm{F}\tau_{\mathrm{ee}}$ numerically according to the results of Jungwirth and MacDonald \cite{jungwirth_electron-electron_1996}, which contains corrections compared to the analytical expression by Giuliani and Quinn \cite{giuliani_lifetime_1982}. The details are described in the following.
We calculate the electron-electron scattering length from the imaginary part of the retarded quasiparticle self-energy $\Sigma(k,\omega)$ evaluated at the Fermi surface, {\it i.e.} $\ell_{\rm ee}^{-1} = -2 \Im m[\Sigma(k_{\rm F}, 0)] /(\hbar v_{\rm F})$. Here $k_{\rm F} = \sqrt{2 \pi n}$ and $v_{\rm F} = \hbar k_{\rm F}/m_{\rm GaAs}$ are, respectively, the Fermi momentum and velocity, whereas $n$ is the electron density and $m_{\rm GaAs} = 0.067 m_{\rm e}$ is the effective electron mass in the GaAs quantum well ($m_{\rm e} = 9.1\times 10^{-31}~{\rm Kg}$ is the bare electron mass). 
The self-energy $\Sigma(k,\omega)$ is calculated within the $G_0W$ approximation,~\cite{giuliani_quantum_2008} {\it i.e.}
\begin{equation}\label{eq:selfenergyg0w}
\Im m[\Sigma(k, \omega)] = \int \frac{d^2{\bm q}}{(2\pi)^2}\Im m[W(q,\omega -  \xi_{{\bm k} - {\bm q}})]~ [ n_{\rm B}(\omega - \xi_{{\bm k} - {\bm q}}) + n_{\rm F}(-\xi_{{\bm k} - {\bm q}})]
~,
\end{equation}
where $n_{\rm F/B}(\varepsilon) = (e^{\beta \varepsilon} \pm 1)^{-1}$ stand for the equilibrium Fermi and Bose distributions, respectively, $\beta = (k_{\rm B} T)^{-1}$ is the inverse temperature ($k_{\rm B}$ is the Boltzmann constant), and $\xi_{{\bm k}} = \hbar^2 k^2/(2 m_{\rm GaAs}) - \varepsilon_{\rm F}$ is the band energy from the Fermi energy $\varepsilon_{\rm F} = \hbar^2 k_{\rm F}^2/(2 m_{\rm GaAs})$. In Eq.~(\ref{eq:selfenergyg0w}) $W(q,\omega) = V(q,\omega)/\epsilon(q,\omega)$ is the screened Coulomb interaction, while $V(q,\omega) = 2\pi e^2/(\epsilon_{\rm GaAs} q)$ is the Fourier transform of the bare Coulomb interaction and $\epsilon(q,\omega) = 1 - V(q,\omega) \chi_{nn}(q,\omega)$. Here $\chi_{nn}(q,\omega)$ is the density-density linear-response function, while $\epsilon_{\rm GaAs} = 12$ is the dielectric constant of GaAs. Note that $\Im m[\Sigma(k, \omega)]$ as defined from Eq.~(\ref{eq:selfenergyg0w}) only depends on the modulus of ${\bm k}$. The dependence on the angle it forms with the ${\hat {\bm x}}$-axis can be removed by a change of variables in the integral on the right-hand side of Eq.~(\ref{eq:selfenergyg0w}).

\begin{figure}
 \includegraphics[width=0.65\linewidth]{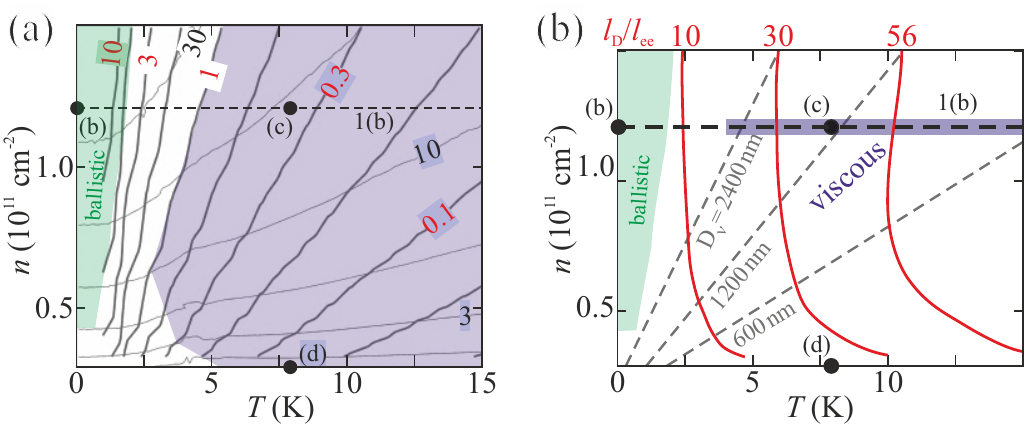}
 \caption{(a) Absolute values of $l_\mathrm{ee}$ and $l_\mathrm{D}$ as a function of $T$ and $n$, from which we extract the ratio $l_\mathrm{D}/l_\mathrm{ee}$. Thick lines with red numbers show calculated $l_\mathrm{ee}/\SI{1}{\micro\meter}$. Thin lines with black numbers denote $l_\mathrm{D}/\SI{1}{\micro\meter}$. The ballistic regime with $l_\mathrm{ee}>\SI{5}{\micro\meter}$, $l_\mathrm{D}>\SI{5}{\micro\meter}$ is shaded green. The blue shade marks the regime of $l_\mathrm{ee}<l_\mathrm{D}/10$ and $l_\mathrm{ee}<\SI{1}{\micro\meter}$. The labels (b), (c), (d) mark the parameters of the SGM measurements in Fig.~2 of the main text, and the dashed horizontal line marks the density of the measurement in Fig.~1(b).
 (b) From the absolute values in (a) we extract the ratio $l_\mathrm{D}/l_\mathrm{ee}$ as a function of $T$ and $n$, which is illustrated by the red contours (repetition of Fig.~2(a) of the main text).
}
 \label{fig_lee_lD_contour}
\end{figure}

Knowing both scattering lengths in the full parameter space spanned by $n$ and $T$, we show a contour plot of $l_{\mathrm{D}}$ and $l_\mathrm{ee}$ in Fig.~\ref{fig_lee_lD_contour}(a). Even though transport regimes do not have abrupt limits, we indicate two areas where ballistic and viscous effects are expected according to the following rules: 
We mark the ballistic regime, where both $l_\mathrm{ee}$ and $l_\mathrm{D}$ exceed \SI{5}{\micro\meter}, by the green shade. Viscous effects are expected if $l_\mathrm{ee}$ is well below both $l_\mathrm{D}$ and the sample size. Therefore we indicate the regime with $l_\mathrm{ee}<l_\mathrm{D}/10$ and $l_\mathrm{ee}<\SI{1}{\micro\meter}$ by the blue shade. The black dots and the dashed line mark the parameters of the measurements in the main text.
Figure~\ref{fig_lee_lD_contour}(b) reproduces Fig.~2(a) for convenient comparison. 

\begin{figure}
 \includegraphics[width=\linewidth]{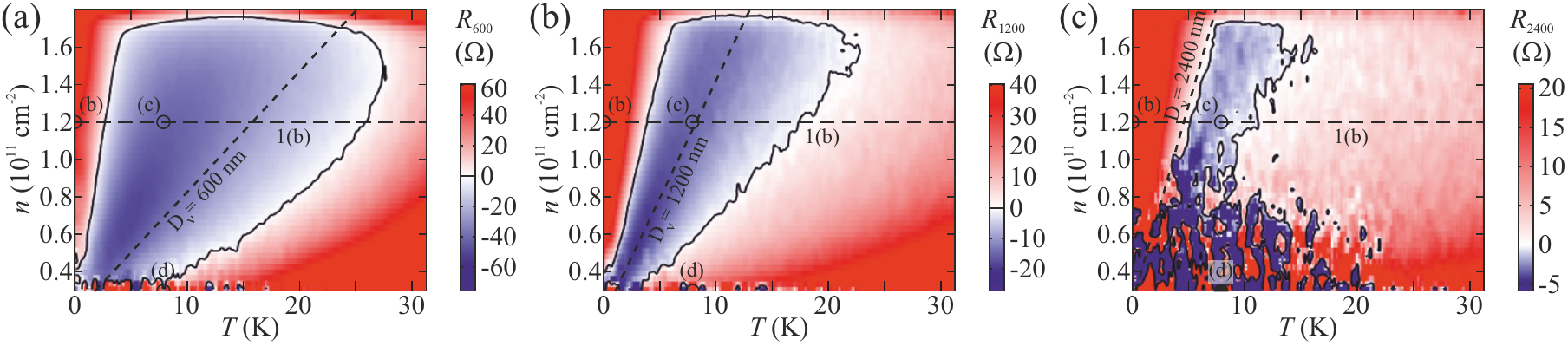}
 \caption{Vicinity resistances as a function of temperature and charge carrier density. The parameters of the measurements in the main text are marked by the circles and the horizontal line. The inclined line marks the situation, when the separation between the current injecting orifice and the voltage probe equals $D_\nu$. These measurements are obtained in a second cool-down of the same sample and details might differ from the other presented data. At low density, the signals show noise of unknown origin, which was not present during the first cool-down. 
}
 \label{fig_Rvic_vs_n_T}
\end{figure}

In a second cool-down of the same sample, we measured the vicinity resistance as a function of the full accessible parameter range of charge carrier density and temperature, but without scanning gate microscopy. The three measured vicinity resistances are shown in Fig.~\ref{fig_Rvic_vs_n_T} as color plots with a black line highlighting the sign inversion. The circles and horizontal line marks the parameters of the measurements in Fig.~1(b) and Figs.~2(b)-(d) of the main text. The inclined dashed lines indicate the parameters where the characteristic length $D_\nu$ of the hydrodynamic effect is equal to the distance between the current injecting opening and the corresponding voltage probe. These lines correspond to the grey dashed lines in Fig.~\ref{fig_lee_lD_contour} and in Fig.~2(a) of the main text.

The results show an extended parameter range of negative vicinity resistances, around the temperature where $D_\nu$ is similar as the separation of the voltage probe to the current injecting orifice. Towards high charge carrier density, the vicinity resistances become positive at a density coinciding with the formation of the second 2DEG as shown above in Fig.~\ref{fig_mobi_dens}. 


\section{Hydrodynamic model with locally reduced electron density}
As the SGM tip induces an approximately Lorentzian shaped potential in the 2DEG \cite{eriksson_effect_1996}, we approximate the electron density at a position $x',y'$ in the channel by 
\begin{equation}
n(x',y')=n_0-1.2 n_0 \frac{l^2}{(x'-x)^2+(y'-y)^2+l^2}
\label{eq_Lorentz}
\end{equation}
with $n_0$ the electron density in absence of the tip,  a FWHM $l=\SI{300}{\nano\meter}$, and a cut-off at zero (depleted 2DEG). The Comsol simulated charge distribution described in the next section supports this model of $n(x',y')$ in the vicinity of the tip.

We simulate the SGM experiment by solving the hydrodynamic model from Bandurin \textit{et al.} \cite{bandurin_negative_2016} for every tip position on a line \SI{0.5}{\micro\meter} from the upper channel boundary. These tip positions correspond to the dashed lines in Fig.~2(c) and (d) if we take a 2DEG depletion length of \SI{150}{nm} around the top-gates into account.

\section{Trajectory calculations}
\begin{figure}
 \includegraphics[width=0.6\linewidth]{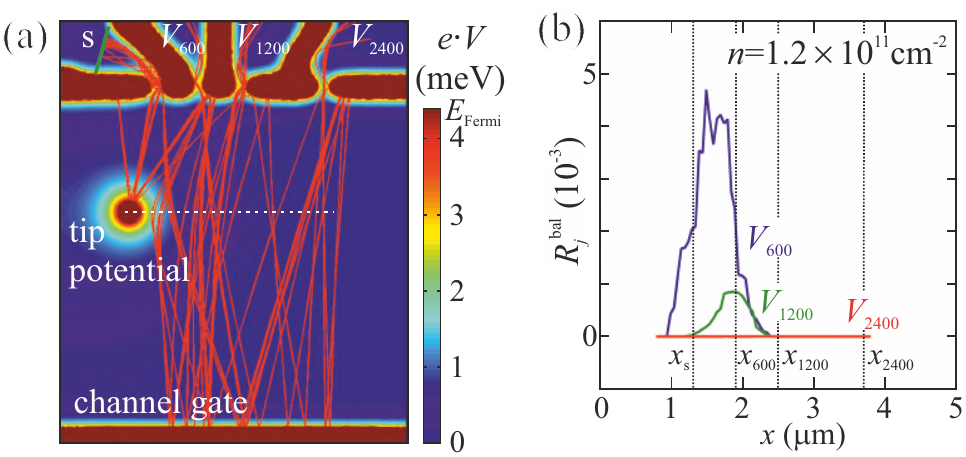}
 \caption{Repetition of Fig.~5 in the main text with labels marking the 2DEG leads to the orifices in (a).
 }
 \label{fig_traj_label}
\end{figure}

We model the potential in the 2DEG caused by SGM tip and QPC gates by calculating the charge distribution in Thomas-Fermi approximation with the finite element software COMSOL 5.0. The sample geometry includes the layer thickness of the Ga[Al]As heterostructure, the SGM tip size and the electron-beam lithography defined top-gates. The resulting electrostatic potential for one tip position is shown as a color plot in Fig.~\ref{fig_traj_label}(a) as well as in Fig.~5(a) of the main text.

Using this potential, we calculate the classical trajectories of electrons at the Fermi energy at an electron density $n=\SI{1.2e11}{\per\square\centi\meter}$. The trajectories start equidistantly and with a homogeneous angle distribution in the source lead, in Fig.~\ref{fig_traj_label}(a) the starting line is indicated in green in the upper left corner. The red lines show post-selected trajectories that end in one of the three vicinity voltage probes $V_{600}$, $V_{1200}$, and $V_{2400}$. So far, no random scattering is included in the calculation. 

We only consider electrons that have not scattered after leaving the source contact. We neglect the contributions of scattered electrons and their scattering partners for the sake of simplicity. The number of electrons that did not experience a scattering event decreases exponentially with trajectory length $l$. We therefore introduce a weight that exponentially decreases with $l$. 
As a qualitative measure $R_j^{\mathrm{bal}}$ of the ballistic contribution to the vicinity resistance $R_j$ we count the weighted number of trajectories ending in the voltage probe $V_j$
$$R_j^{\mathrm{bal}}=\mathop{\sum_{\mathrm{trajectory} \, k }}_{\mathrm{ \, ends \, in} \, V_j} e^{-l_k / l_{s}}$$
with $l_{s}$ the typical length scale of scattering. At the parameters of Fig.~4(a) in the main text ($T=\SI{7.9}{\kelvin}$, $n=1.2\times 10^{11} \, \mathrm{cm}^{-2}$), electron-electron scattering is dominant ($l_\mathrm{ee}\approx \SI{370}{\nano\meter} \ll l_\mathrm{D} \approx \SI{15}{\micro\meter}$) so we use $l_{s}=\SI{400}{\nano\meter}$. 
Despite of its simplicity, this qualitative trajectory simulation illustrates the tip-position dependence of ballistic effects. We find a maximum of $R_\mathrm{vic}$ if the tip is in $x$-direction in the middle between the source orifice and the respective voltage probe, which agrees with the experimental results at high density described in the main text.

This model neglects the contribution of the scattered electrons because it is beyond our capabilities to calculate. With the assumption, that the contribution of the scattered electrons is independent of tip position, we expect a vertical shift of the results in Fig.~\ref{fig_traj_label}(b), which does not influence the $x$-position of the maxima.

\end{document}